\documentclass[aip,pof,showpacs,showkeys,nofootinbib]{revtex4}
\usepackage{graphicx}
\usepackage{amssymb}
\usepackage{amsmath}
\usepackage{euscript}
\usepackage{esint}
\usepackage{color}

\usepackage{amsfonts,amsthm}           % Basic AMSTeX packages
\usepackage{dsfont}                                    % %Double-stroke font
\newcommand{\pd}{\partial}                % partial differential/derivative
\newcommand{\ii}{\mathrm{i}}              % imaginary unit
\newcommand{\diff}{\mathrm{d}}            % (full) differential
\newcommand{\bigO}{\mathit{O}}            % Landau's order-of    %   \newcommand{\bigO}{\mathcal{O}}
\newcommand{\bvec}[1]{{\boldsymbol{#1}}}  % 3D vector
\newcommand{\Reals}{\mathds{R}}           % real set
\newcommand{\Hilbert}{\EuScript{H}}       % 'Curly H'
\DeclareMathOperator{\IIm}{\mathrm{Im}}   % imaginary part
\DeclareMathOperator{\sgn}{\mathrm{sign}} % signum

\newcommand{\vsp}{\vspace{0.5em}}        % a small vertical indent

\newcommand{\etc}{\textit{etc.}}         % et cetera
\newcommand{\eg}{\textit{e.g.}}          % exempli gratia

\begin{document}

\title{
    Numerical study of strongly-nonlinear regimes of steady premixed flame propagation. \\
    The effect of thermal gas expansion and finite-front-thickness effects
}

\author{Kirill A.~Kazakov}

\author{Oleg G.~Kharlanov}

\affiliation{Department of Theoretical Physics, Faculty of Physics, Lomonosov Moscow State University, 119991, Moscow, Russian Federation}

\begin{abstract}
    Steady propagation of premixed flames in straight channels is studied numerically using the on-shell approach. A first
    numerical algorithm for solving the system of nonlinear integro-differential on-shell equations is presented.
    It is based on fixed-point iterations and uses simple (Picard) iterations or the Anderson acceleration method that facilitates separation of
    different solutions. Using these techniques, we scan the parameter space of the problem so as to study various effects
    governing formation of curved flames. These include the thermal gas expansion and the finite-front-thickness effects,
    namely, the flame stretch, curvature, and compression. In particular, the flame compression is demonstrated to have a
    profound influence on the flame, strongly affecting the dependence of its propagation speed on the channel width $b$.
    Specifically, the solutions found exhibit a sharp increase of the flame speed with the channel width. Under a weak flame
    compression, this increase commences at $b/\lambda_{\text{c}}\approx 2 \div 3$, where $\lambda_{\text{c}}$ is the cutoff wavelength,
    but this ratio becomes significantly larger as the flame compression grows. The results obtained are also used to identify limitations of the analytical approach based on the weak-nonlinearity assumption, and to revise the role of noise in the flame evolution.
\end{abstract}

% PACS codes
% 47.20.-k    Flow instabilities
% 47.32.-y    Vortex dynamics; rotating fluids
% 47.70.Pq    Reactive and radiative flows >>> Flames; combustion
% 82.33.Vx    Reactions in various media >>> Reactions in flames, combustion and explosion
% 02.60.-x    Numerical approximation and analysis
% 02.60.Nm    Numerical approximation and analysis >>> Integral and integrodifferential equations
% 02.60.Cb    Numerical approximation and analysis >>> Numerical simulation; solution of equations
%


\maketitle

\section{Introduction}\label{sec:Introduction}%
Flame propagation in gaseous mixtures is a phenomenon that involves several processes of quite different nature, which are
characterized by quite different time and length scales, and yet are so much interrelated by various strongly nonlinear
interactions that altering one of them can entirely change the whole picture. The heat and mass transport inside the flame
front govern the evolution of the short-wavelength flame perturbations that rapidly grow and coalesce to form larger patterns,
whose structure depends on the global conditions (geometry of the combustion domain, gravity, incoming flow vorticity, \etc);
formation of the latter, in turn, induces large-scale flows that affect the transport inside the flame front. An accurate
account of all these processes is therefore necessary for a quantitative description of the flame dynamics. Because of an
extremely large scale separation, direct numerical simulations (DNS) are not very helpful in this respect, as their
computational demands for typical laboratory conditions largely exceed the computational resources that are presently
available. A great deal of analytical work is thus needed to reduce the system of governing equations so that it could be
efficiently solved numerically.

The first step in this direction is to consider the flame as a surface of discontinuity in the physicochemical properties of
the gases. In fact, the flame front thickness is normally several orders of magnitude smaller than the global length scales.
Also, since the process is essentially subsonic, with a great accuracy, the gas flows can be considered incompressible. Thus, the gas densities in the upstream and downstream regions of
the discontinuity surface are assumed to be uniform quite up to this surface, and to coincide with the bulk density of fresh
and burnt gases, respectively. The effects of transport processes inside the flame front manifest themselves in this picture in
the form of jump conditions for the flow variables at the discontinuity surface (which for brevity is also called the front),
and in the expression for its normal speed \cite{mikhelson1890, markstein1951, markstein1964, matalon1982, pelce1982,
clavin1985}. Explicit calculation of these finite-front-thickness effects is rather laborious for real flames, but their
general structure can be established without much effort by identifying possible geometrical invariants of an appropriate
differential order. In particular, in the first order with respect to the flame front thickness (which is sufficient for most
applications), these include the so-called flame stretch and flame curvature. The finite-front-thickness corrections appear in
the expressions for the gas velocity and pressure jumps with three independent parameters termed collectively the Markstein
lengths. Therefore, all interactions of the outer gas flows with the processes inside the flame front are properly taken into
account once the values of these parameters in a given mixture are specified. They can be either measured experimentally, or
inferred from DNS of simple configurations such as spherical flames.

The next step is a partial integration of the equations that govern the propagation of the front separating two
constant-density fluids, with the aim of reducing them to equations for the flow variables restricted to the front. When
successful, this reduction yields the flame front dynamics formulated in inner terms, that is, as a system of equations for
functions defined on the front. Implementation of this programme is relatively easy under the assumption of weak flow
nonlinearity. Namely, a single equation for the front position has been obtained that describes the linear evolution of
perturbations of planar flames \cite{darrieus, landau, markstein1964}. However, a problem with the weak-nonlinearity assumption
is that it is justified only within an initial time interval of the order $\lambda_{\text{c}}/U_{\text{f}}$, where
$\lambda_{\text{c}}$ is the short wavelength cutoff of unstable perturbations and $U_{\text{f}}$ is the planar-front speed
relative to the fresh gas. In practice, this time is typically measured in milliseconds. At later times, the flame can remain
weakly curved only if the gas expansion coefficient $\theta$ (the fresh-to-burnt gas density ratio) is close to unity, that is,
when the density contrast across the front is relatively small \cite{siv, sivclav}. At the same time, for most flames of
interest, $\theta$ is much greater than unity (it is typically from 5 to 8 for flames in strongly diluted mixtures such as
hydrocarbon-air, while for near-stoichiometric methane-oxygen mixtures, $\theta \approx 10$).

In the general nonlinear case (arbitrary $\theta$), reduction to a system of equations for the flow variables restricted to the
front (briefly, for their \textit{on-shell} values) has been accomplished for two-dimensional flames. This so-called on-shell
description was first developed in the case of steady flame propagation in straight channels \cite{kazakov1} and then extended
to unsteady flames \cite{jerk1} in channels of varying width \cite{jerk2, note1}. It enabled analytical study of extremely
nonlinear propagation regimes characterized by strong flame elongation, such as that of flames anchored in high-velocity
streams \cite{kazakov2}, gravity-driven flame propagation in horizontal tubes \cite{kazakov3}, and near-extinction phenomena in
vertical tubes \cite{kazakov4, kazakov5}.

Yet, with the exception of the above limiting cases, complexity of the on-shell equations does not permit further analytical
advancement, and in intermediate situations like the spontaneous cell formation or the flame evolution in a moderate gravity,
these equations need to be solved numerically. The purpose of the present paper is to present a numerical method we have
recently developed to study the on-shell equations, and to apply it to steady flame propagation in straight channels. Various
steady regimes will be identified, and a comparison with the earlier results existing in the literature will be made. Finally, we
will use our results to settle several open issues of premixed flame propagation.

The paper is organized as follows. In Sec.~\ref{sec:Equations}, we display the system of integro-differential equations to be
solved. The method for solving these equations via fixed-point iterations is described in Sec.~\ref{sec:Algorithm}. Two
approaches are presented -- based on simple (Picard) iterations and on the Anderson acceleration method (see
Ref.~\cite{Anderson} and, \eg, Refs.~\cite{Olshansky_IterativeMethods, Matveev_Anderson} for modern applications, including
determination of steady solutions of nonlinear integral equations). A large number of numerical solutions obtained are given in
Sec.~\ref{sec:Results}, where they are compared with the results already known and used to study in detail the effect of gas
expansion and of the finite-front-thickness effects on the flame structure. Conclusions are drawn in Sec.~\ref{sec:Conclusion}.
The paper has an Appendix where an equation for the flame front position is derived in the first post-Sivashinsky
approximation taking into account the flame compression effect.

\section{The flame model and the on-shell equations}\label{sec:Equations}

\begin{figure}[ht]
    \includegraphics[width=7cm]{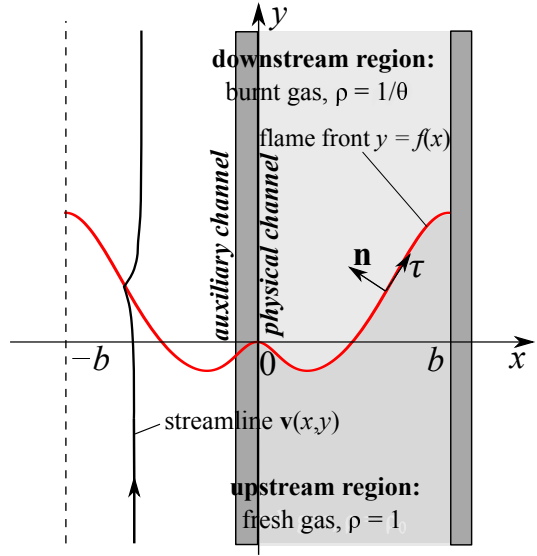}
    \caption{Geometry of steady flame propagation in a straight channel in the rest frame of the flame front}
    \label{fig1}
\end{figure}

Consider a 2D steady flame propagating in an initially quiescent gaseous mixture filling a straight channel of width $b$ with
ideal walls. We choose Cartesian coordinates $(x,y)$ so that the $y$-axis coincides with one of the channel walls, $x \in
(0,b)$ being the channel interior and $y \to -\infty$ being in the fresh gas (Fig.~\ref{fig1}). The $x$- and $y$-
components of the gas velocity field $\bvec{v}(x,y)$ will be denoted $w(x,y)$ and $u(x,y)$, respectively. On neglecting the
heat losses, the flow incompressibility means that the gas density is constant except in a narrow region near the reaction
zone, which is characterized by large density gradients. The hydrodynamic flame model replaces this region (whose
characteristic thickness is called the flame front thickness $l_{\text{f}}$) with a surface of discontinuity and assumes that
the gas density upstream and downstream of this surface is uniform quite up to the surface, coinciding with the bulk density of
fresh and burnt gases, respectively. This discontinuity surface is called the flame front and its exact location with respect
to the original large density-gradient region it replaces is unambiguously fixed by the requirement that the positions of gas
elements which are remote from the flame front at any given instant be unaffected by this replacement \cite{kazakov5}. In the
rest frame of the flame, the front position can be written as $y = f(x),$ with the origin of the coordinate system chosen so
that $f(0) = 0.$ Natural units, in which $b = U_{\text{f}} = 1$, will be used throughout. Here, $U_{\text{f}}$ is the planar flame speed relative to the fresh gas, which is also the normal
fresh gas velocity at the steady front in the zero-order approximation with respect to $l_{\text{f}}$. The fresh gas density
taken to be unity, that of the burnt gas is $1 / \theta$.

Next, we assume that the flame propagation produces no flow separation at the front endpoints (there is no stagnation in the
burnt gas flow). Together with the wall impermeability, this implies the following boundary conditions
\begin{equation}
    w(0,y) = w(1,y) = f'(0) = f'(1) = 0,
\end{equation}
where prime denotes a derivative with respect to $x$. The on-shell equations take their simplest form under these conditions,
because the influence of the channel walls can be simply taken into account by considering the channel flow as a part of an
unbounded flow filling the whole $xy$ plane, which is obtained by a periodic continuation of the given flame pattern along the
$x$-axis. First, it is mirrored onto an auxiliary channel $x \in [-1,0]$ according to
\begin{equation}\label{reflect}
    f(x) = f(-x),\quad w(x,y) = -w(-x,y), \quad u(x,y) = u(-x,y).
\end{equation}
After that, the flow in the extended channel $x \in [-1,1]$ is periodically continued along the whole $x$-axis. Then the
on-shell fresh gas velocity satisfies the following complex integro-differential equation (the master equation)
\cite{kazakov1,jerk2}
\begin{equation}\label{masterEquation}
    2\omega_-' + \left(1 +
    \ii\hat\Hilbert\right)\left\{[\omega]' -
    \frac{Nv^n_+\sigma_+\omega_+}{v^2_+} + \frac{1 + \ii f'}{2}\int_{-1}^{+1}\diff\xi\; \frac{Nv^n_+\sigma_+\omega_+(\xi)}{v^2_+(\xi)}\right\} = 0,
\end{equation}
where $\omega = u + \ii w$ is the complex velocity and $[\omega] = \omega_+ - \omega_-$ is its jump across the front; the
$-(+)$ subscript denotes restriction to the front of a function defined upstream (downstream) of the front, \eg, $w_-(x) = w(x,
f(x) - 0)$; $N = \sqrt{1 + f'^2}$; $v^n_\pm = \bvec{v}_\pm \cdot \bvec{n}$ is the normal gas velocity ($\bvec{n}$ being the
unit vector normal to the front pointing towards the burnt gas, see Fig.~\ref{fig1}), $\sigma = \pd u/\pd x - \pd
w / \pd y$ is the gas flow vorticity; finally, the operator $\hat{\Hilbert}$ is defined on $2$-periodic functions with zero
mean across the channel as
\begin{equation}\label{hcurvedf}
    \left(\hat{\Hilbert}a\right)(x) = \frac{1 + \ii f'(x)}{2}~\fint_{-1}^{+1}\diff\xi
    \cot\left\{\frac{\pi}{2}(\xi - x + \ii[f(\xi) - f(x)])\right\} \; a(\xi),
\end{equation}
with the slash denoting the principal value of the integral.

To complete the system of equations for the three unknown functions $u_-(x), w_-(x), f(x)$, one has to specify expressions for
the normal gas velocity, velocity jumps, and $\sigma_+.$ The first one is the so-called evolution equation
\begin{equation}\label{evolutionEquation}
    v_-^n = 1 - S(f', \omega_-),
\end{equation}
where $S(f', \omega_-)$ is a quasilocal functional of its arguments that describes the finite-front-thickness effects (the
complex velocity is used as an argument of the real $S$ only for brevity; $S$ depends separately on $u_-$ and $w_-,$ as well as
on their derivatives). Within the first order in the flame front thickness, it reads
\begin{equation}\label{evolutionEquation_S}
    S = \mathcal{L}_{\text{c}} \frac{f''}{N^3} + \mathcal{L}_{\text{s}} \frac{(v_-^\tau)'}{N},
\end{equation}
where $\mathcal{L}_{\text{c}, \text{s}} = O(l_{\text{f}})$ are the parameters (Markstein lengths) quantifying the effects of
the flame curvature and its stretch on the local burning rate; $v_-^\tau \equiv \bvec{v}_- \cdot \bvec\tau = (w_- + f' u_-)/N$
is the tangential component of the on-shell fresh gas velocity.

Next, the gas velocity jumps follow from the equations (as proved in Ref.~\cite{kazakov5}, these familiar zero-order relations
hold true on inclusion of the first-order finite-front-thickness corrections)
\begin{equation}
    v_+^\tau = v_-^\tau, \quad v_+^n = \theta v_-^n
\end{equation}
and can be summarized in the expression for the complex velocity jump
\begin{equation}\label{omegaJump}
    [\omega] = (\theta - 1)\frac{1 - \ii f'}{N}v^n_-.
\end{equation}
Finally, the value of the burnt gas vorticity at the front can be found from the gas pressure jump \cite{hayes1957, kazakov5}
\begin{equation}\label{sigmaJump}
    \sigma_+ = [\sigma] = -\frac{1}{N v_-^n}\left\{ \frac{\theta - 1}{2 \theta} (v^\tau_-)^2
                        + \frac{\theta - 1}{2} (v^n_-)^2 + 2 \mathcal{L}_\sigma
                            \frac{f''}{N^3}
                        \right\}',
\end{equation}
wherein $\sigma_- = 0$ by the Thomson theorem. The  constant $\mathcal{L}_\sigma$ entering the above expression is the last of
the three independent lengths parameterizing the first-order finite-front-thickness effects. The corresponding contribution to
$\sigma_+$ (or to the pressure jump) describes the so-called flame compression effect (which has nothing to do with the gas
compressibility). This term was coined in Ref.~\cite{class2003}, where importance of the effect with regard to the unstable
flame evolution was also alluded. As we show below, it strongly impacts the steady flame structure indeed. $\mathcal{L}_\sigma$
is always positive; it vanishes for $\theta \to 1,$ but rapidly grows with $\theta,$ and for flames of practical interest
(large $\theta$ and realistic temperature dependence of the kinetic coefficients), $\mathcal{L}_\sigma \sim
l_{\text{f}}\theta^{3/2}$.

\section{The numerical scheme}\label{sec:Algorithm}

The system in question consists of one complex Eq.~\eqref{masterEquation} [complemented by the jump conditions
\eqref{omegaJump}, \eqref{sigmaJump}] and one real Eq.~\eqref{evolutionEquation}. The three unknowns are the $2$-periodic
functions $w_-(x), u_-(x)$, and $f(x),$ obeying reflection conditions \eqref{reflect}. It proves useful to combine the real
number $u_-(0) \equiv u_0,$ which in what follows will be refereed to as the edge velocity, and the front position into a
single (real even) function
\begin{equation}\label{phi_def}
    \varphi(x) = f(x) + u_0.
\end{equation}
The choice $f(0) = 0$ implies that $u_0, f$ can be expressed via $\varphi$ as $u_0 = \varphi(0)$, $f(x) = \varphi(x) -
\varphi(0)$.

The iterative procedure we use for numerical solving of the system is as follows. Consider an approximation $\varphi^{(p)}(x)$
of $\varphi^\star(x),$ which is part of the solution $\bigl(w_-^\star(x),u_-^\star(x),\varphi^\star(x)\bigr)$ of the system.
With a fixed $\varphi = \varphi^{(p)}$, the master equation~\eqref{masterEquation} is a nonlinear integro-differential equation
for the function $\omega_-(x)$. Its solutions can be found as fixed points of a recurrence relation
\begin{gather}\label{masterEquation_implicit}
    \omega_-^{(m+1)}(x) = u_0 + \int_0^x \diff\xi \; K[\omega_-^{(m)}, \varphi](\xi), \qquad m = 0, 1, 2,  \ldots,
\end{gather}
where $K$ is the nonlinear integro-differential operator determining $\omega_-'$ according to Eq.~\eqref{masterEquation}.
Actual simulations show that iterations (\ref{masterEquation_implicit}) rapidly converge to a unique solution
$\omega_-^\star[\varphi]$ for any reasonable function $\varphi(x)$ and iteration seed $\omega_-^{(0)}(x)$. After that, to
construct a new approximation $\varphi^{(p+1)}(x)$ of $\varphi^\star(x)$ (that is, a new $f^{(p+1)}(x)$ and a new
$u^{(p+1)}_0$), we use the evolution equation~\eqref{evolutionEquation} written as
\begin{equation}\label{evolutionEquation_recurrence}
    v_-^{n(p+1)} = 1 - S\bigl(f^{(p+1)}{}', \; \omega_-^\star[\varphi^{(p)}] + u_0^{(p+1)} - u_0^{(p)}\bigr),
\end{equation}
with the left-hand side understood as
\begin{equation}\label{evolutionEquation_recurrence_lhs}
    v_-^{n(p+1)} = \frac{u_-^\star[\varphi^{(p)}] + u_0^{(p+1)} - u_0^{(p)} - f^{(p+1)}{}' w_-^\star[\varphi^{(p)}]}{\sqrt{1 + f^{(p+1)}{}'\,^2}}.
\end{equation}
Note that $\omega_-^\star[\varphi^{(p)}]$ enters the above equations in a `corrected' form: it is shifted by $(u_0^{(p+1)} -
u_0^{(p)})$ in order to match the edge velocity of the next (unknown) iteration. $f^{(p+1)}(x)$ is the solution of the ordinary
differential equation (\ref{evolutionEquation_recurrence}). Together with the boundary conditions at the channel walls, this
equation constitutes an overdetermined system, and therefore, implies a condition on the iterated edge velocity $u_0^{(p+1)}$.
In fact, a \textit{first}-order ordinary differential equation \eqref{evolutionEquation_recurrence} in its unknown
$f^{(p+1)}{}'(x)$ is to be solved with \textit{two} boundary conditions $f^{(p+1)}{}'(0) = f^{(p+1)}{}'(1) = 0$. Such a system
may have solutions only for special values of $u_{0}^{(p+1)}$. Indeed, if we integrate it with the initial condition
$f^{(p+1)}{}'(0) = 0$, then the other condition yields an algebraic equation for $u_{0}^{(p+1)}$
\begin{equation}\label{evolutionEquation_u0Equation}
    f^{(p+1)}{}'[\omega_-^\star[\varphi^{(p)}], u_{0}^{(p+1)}](1) = 0.
\end{equation}
Thus, in this computational scheme, $u_0$ plays the role of eigenvalue. Its approximation $u_{0}^{(p+1)}$ is found by
scanning a vicinity of $u_{0}^{(p)},$ that is, by solving Eq.~\eqref{evolutionEquation_recurrence} with different
$u_{0}^{(p+1)}$ until condition \eqref{evolutionEquation_u0Equation} is met (a technique for solving boundary-value problems
usually referred to as the shooting method). Note also that for actual flames, shooting can be done efficiently using
bisections, since the left-hand side of Eq.~\eqref{evolutionEquation_u0Equation} turns out to be a monotonic function of
$u_{0}^{(p+1)}$. After the latter is found, the corresponding $\varphi^{(p+1)}$ can be evaluated as
\begin{equation}\label{I_phi}
    \varphi^{(p+1)}(x) = u_{0}^{(p+1)} + \int_0^x \diff\xi \; f^{(p+1)}{}'[\omega_-^\star[\varphi^{(p)}], u_{0}^{(p+1)}](\xi) \equiv
                        \mathcal{I}[\varphi^{(p)}](x).
\end{equation}
We conclude that if the described iterative procedure converges, the sought-after solution $\bigl(
w_-^\star,u_-^\star,\varphi^\star \bigr)$ can be found as
$\bigl(w_-^\star[\varphi^\star],u_-^\star[\varphi^\star],\varphi^\star \bigr)$, where $\varphi^\star$ is a fixed point of the
map \eqref{I_phi},
\begin{equation}\label{evolutionEquation_fEquation}
    \mathcal{I}[\varphi^\star] = \varphi^\star.
\end{equation}

\vsp

The fixed point $\varphi^\star$ does not necessarily have to be reached using simple iterations (the Picard method),
\textit{i.e.}, by repeated application of $\mathcal{I}$ to some seed $\varphi^{(0)}$. For instance, one can introduce a weight
$\beta \in (0, 1]$ to redefine the iteration as
\begin{equation}\label{Picard_iteration_beta}
    \varphi_{p+1} = (1 - \beta) \varphi_p + \beta \mathcal{I}[\varphi_p],
\end{equation}
so that $\varphi^\star$ is also a fixed point of the modified recurrence relation%
\footnote{To avoid confusion with Picard iterates $\varphi^{(p)},$ we use a subscript notation $\varphi_p$ for non-Picard
iterates. Later on, subscripts will also be used for Anderson iterates.}.
As is well known, both conventional and weighted
iterations converge in a sufficiently small vicinity of the fixed point, provided that $\mathcal{I}$ is a contraction at this
point, \textit{i.e.}, $\| \mathcal{I}[\varphi] - \varphi^\star \| < \lambda \| \varphi - \varphi^\star\|$ for some $\lambda \in
(0, 1)$ and $\varphi \to \varphi^\star$ \cite{Olshansky_IterativeMethods}. In the present case, however, this condition turns
out to be a too severe limitation, because not all of the fixed points happen to correspond to contracting operators
$\mathcal{I}$.

There exist a number of techniques to overcome this difficulty. A notable one is the Anderson acceleration method
\cite{Anderson}, which is virtually a generalization of the Krylov subspace approaches to nonlinear equations
\cite{Olshansky_IterativeMethods}. In this method, the $(p+1)$th approximation is constructed as a weighted average of $d \ge
1$ preceding iterates
\begin{equation}\label{Anderson_iteration}
    \varphi_{p+1} = \alpha_{p,1} \varphi_p + \alpha_{p,2} \varphi_{p-1} + \ldots
                    + \alpha_{p,d} \varphi_{p-d+1}, \qquad \alpha_{p,k} \in \Reals, \qquad
                    \sum\limits_{k=1}^d \alpha_{p, k} = 1,
\end{equation}
the weights $\alpha_{p,1},\ldots,\alpha_{p,d}$ being chosen so as to minimize the Euclidean norm of the residual
$\epsilon_{p+1} = \mathcal{I}[\varphi_{p+1}] - \varphi_{p+1}$. Under the assumption that the iterates are sufficiently close to
the fixed point $\varphi^\star$ [and hence so is $\varphi_{p+1}$, by virtue of Eq.~\eqref{Anderson_iteration}], $\mathcal{I}$
can be linearized, and the residual norm squared can be written as a quadratic form
\begin{eqnarray}
    \epsilon_{p+1} &\equiv& \mathcal{I}[\varphi_{p+1}] - \varphi_{p+1}
                \approx \mathcal{I}[\varphi^\star] + \mathcal{I}'[\varphi^\star](\varphi_{p+1} - \varphi^\star) - \varphi_{p+1}
                \equiv \mathcal{J}\cdot(\varphi_{p+1} - \varphi^\star), \quad \mathcal{J} \equiv \mathcal{I}'[\varphi^\star] - 1,\\
                        \|\epsilon_{p+1}\|^2 &\approx& \| \mathcal{J}(\varphi_{p+1} - \varphi^\star)\|^2
                = \left\|\sum_{k=1}^d \alpha_{p,k} \mathcal{J}(\varphi_{p-k+1}-\varphi^\star)\right\|^2
                = \sum_{k,l=1}^d g_{kl}^{(p)} \alpha_{p,k} \alpha_{p,l}. \label{Anderson_quadraticForm}
\end{eqnarray}

The coefficients of this quadratic form can be evaluated by computing the scalar products of the residuals, $g_{kl}^{(p)} =
(\epsilon_{p-k+1}, \epsilon_{p-l+1})$. Minimization of the residual then reduces to the well-known linear least-squares (LLS)
problem, \textit{i.e.}, to minimization of the quadratic form \eqref{Anderson_quadraticForm} with respect to its arguments
$\alpha_{p,1}, \ldots,\alpha_{p,d-1}$, the last argument $\alpha_{p, d} = 1 - \sum_{k=1}^{d-1}\alpha_{p,k}$ being a linear
function of the others (see, \eg, \cite{LeastSquares}). Not only does the Anderson method permit searching for non-contracting
fixed points, it is also more robust than simple iterations: if, for some reason, the $p$th iteration step has thrown
$\varphi_{p+1}$ away of the fixed point, the residual minimization will suppress the weight of the `bad solution'
$\varphi_{p+1}$ in subsequent iterates, so that one `bad try' will not spoil the whole iterative sequence. Moreover, linear
constraints on $\varphi$ are readily incorporated into the Anderson method. For example, the edge velocity constraint $\xi_-
\le u_0 \equiv \varphi(0) \le \xi_+$ leads to a linear inequality-constrained least-squared (CLS) problem on the weights
$\alpha_{p,k}$, which admits an explicit solution similar to the one for LLS \cite{LeastSquares}. In our simulations, we use
this constraint for separating different solution branches.

It is worth mentioning that the slowest part of the computation, namely, the integral transform \eqref{hcurvedf} requiring
$\bigO(N^2)$ operations for $N$ grid points across the channel, is readily parallelized, since the values of the transform at
different points can be evaluated independently. To give an idea of the efficiency of the algorithm, we mention that most
of the results presented below were obtained on a dual-core Intel Atom D525 1.7GHz machine, where it takes about one minute to
find a solution for a grid with $N = 1000$ and flame slope tolerance of $5 \times 10^{-3}$; we resorted to single-node
multiprocessor computation on a supercomputer only for large series of solutions (about 250) needed to build a three-dimensional plot. Further technical details of the numerical approach implemented in our solver, including
evaluation of the singular-kernel integral transform \eqref{hcurvedf}, the use of non-uniform grids, regularization of the inverse
matrices $g^{(p)\,{-1}}$ necessary for the CLS/LLS minimizations, \etc, will be given elsewhere \cite{OKKK_FlameSolver}.

\section{The applications}\label{sec:Results}

The developed method will now be used to study various effects that control formation of steady flames. In contrast to DNS or
real experiments where all the parameters $\theta,$ $\mathcal{L}_{\text{c},\text{s},\sigma}$ have fixed values specific to the
given mixture, our approach allows separate investigation of the effects by varying one of the parameters while keeping the
others fixed. We first set $\mathcal{L}_{\text{s}} = \mathcal{L}_{\sigma}=0$ to consider the most common mechanism of flame
stabilization by the front curvature effect at several representative values of $\theta.$ This helps to illustrate critical
importance of the gas expansion parameter which is often depreciated in the theoretical combustion. Nonzero
$\mathcal{L}_{\text{s}}, \mathcal{L}_{\sigma}$ are then included to investigate the flame stretch and compression, to infer an
interesting interplay between different effects, and to establish an important universality of the dependence of the flame
speed on the channel width. In general, there exist several solutions for each point
$(\theta,\mathcal{L}_{\text{c}}, \mathcal{L}_{\text{s}}, \mathcal{L}_{\sigma})$ of the parameter space, but for
the sake of brevity, we present only the one maximizing the flame propagation speed $U.$ Though proved only within the
weakly-nonlinear theory \cite{vaynblat}, it seems to be in the nature of free flame evolution that only such solutions can be
stable against small perturbations.

\subsection{The effect of finite gas expansion. A practical limitation of the weak-nonlinearity analysis}\label{gasexpansion}

\begin{figure}[ht]
    \includegraphics[width=9cm]{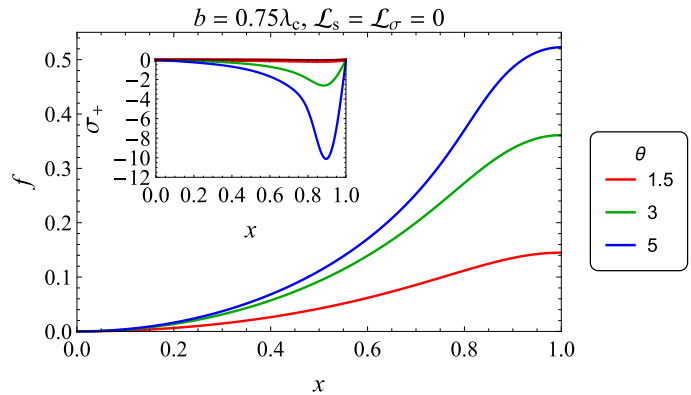}
    \caption{Front position and burnt gas vorticity distribution at the front (inset) for flames with $\theta = 1.5,
    \mathcal{L}_{\text{c}} = 0.036$ (red lines), $\theta = 3, \mathcal{L}_{\text{c}} = 0.072$ (green lines), and $\theta = 5,
    \mathcal{L}_{\text{c}} = 0.086$ (blue lines); channel width $b = 0.75\lambda_{\text{c}}$ in all cases.}
    \label{fig2}
\end{figure}

Steady flame formation in a channel can be described as a nonlinear stabilization of a finite number of unstable modes, which
grow and gradually coalesce into larger and more slowly evolving patterns. In the regime of saturated nonlinearity, the front
slope is not small over most part of the flame unless $(\theta-1)\equiv \alpha$ is small. But since it drops down to zero near the channel walls (by virtue of the boundary conditions existing thereat), the walls appear to exert a stabilizing effect on the flame. This gives hope that the weakly-nonlinear theory could be applicable even to flames with non-small $\alpha$ in
sufficiently narrow channels. However, the downside of this stabilization is a non-uniformity of the gas velocity distribution
across the channel. Namely, to meet the conditions $w=0,$ $f'=0,$ the flow variables have to rapidly vary near the wall (this
happens normally at the trailing edge of the flame), which results in an enhanced vorticity production in this region. This is
illustrated by Fig.~\ref{fig2} showing the front shapes and on-shell vorticities of the burnt gases for flames with
$\theta=1.5, 3$, and $5$ propagating in a channel of width $b=0.75\lambda_{\text{c}}.$ We observe that despite the seemingly
smooth front shapes, vorticity is sharply peaked near the wall $x=b,$ significantly exceeding unity (in natural units) already
for such a moderate gas expansion as $\theta=3.$ At the same time, the small-$(\theta-1)$ expansion implies smallness of the
burnt gas vorticity: once the flame-induced gas velocity variations $\Delta u, \Delta w$ are assumed to be much less than
unity, so must be their derivatives, because $\lambda_{\text{c}}$ and $b$ are the only characteristic lengths in this approach, and
in the case $\lambda_{\text{c}} \simeq b = 1$ one has, {\it e.g.}, $u_+' \simeq \Delta u/\lambda_{\text{c}} \simeq \Delta u \ll
1$ (in fact, the small-$(\theta-1)$ expansion assumes that differentiation of a flow variable raises its smallness order by
one; the key points of this expansion are summarized in the Appendix). We thus see that the assumption of weak vorticity
production is not valid for flames with $\alpha \gtrsim 1$ even in narrow channels. As is evident from the figure, an underlying reason is that for such flames $\lambda_\text{c}$ is not actually the smallest characteristic length: the burnt gas vorticity varies on a much smaller scale $\approx \mathcal{L}_{\text{c}}$.

\begin{figure}[ht]
    \includegraphics[width=9cm]{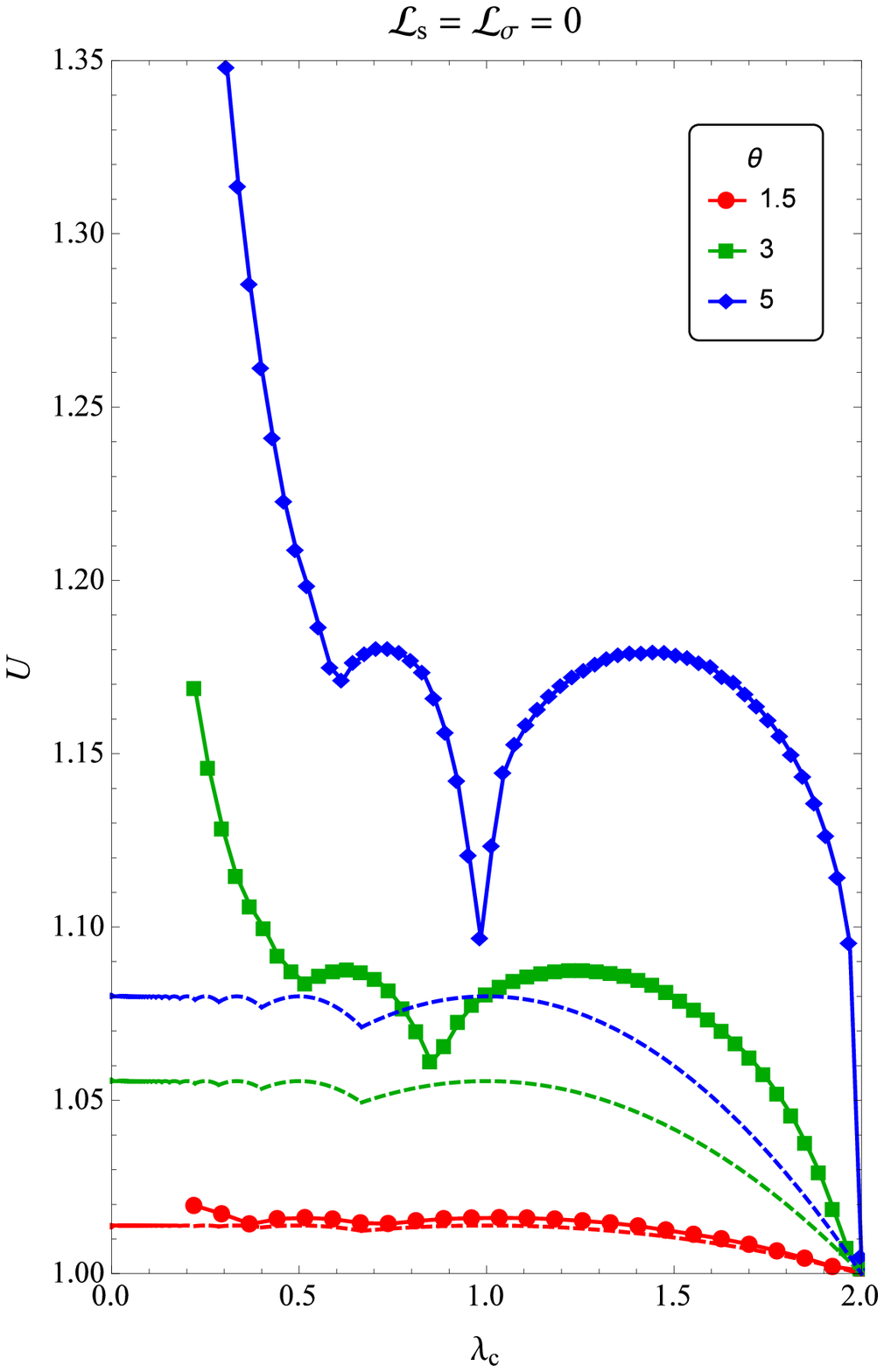}
    \caption{Flame speed versus the cutoff wavelength
    for flames with $\theta = 1.5, 3, \text{ and } 5$ (solid red, green, and blue lines, respectively). $\mathcal{L}_{\text{s}} = \mathcal{L}_{\sigma} = 0$
    in all cases. The corresponding predictions of the weakly-nonlinear theory (solutions of the Sivashinsky-Clavin equation \cite{sivclav}) are shown by
    dashed lines.}
    \label{fig3}
\end{figure}

\begin{figure}[ht]
    \includegraphics[width=9cm]{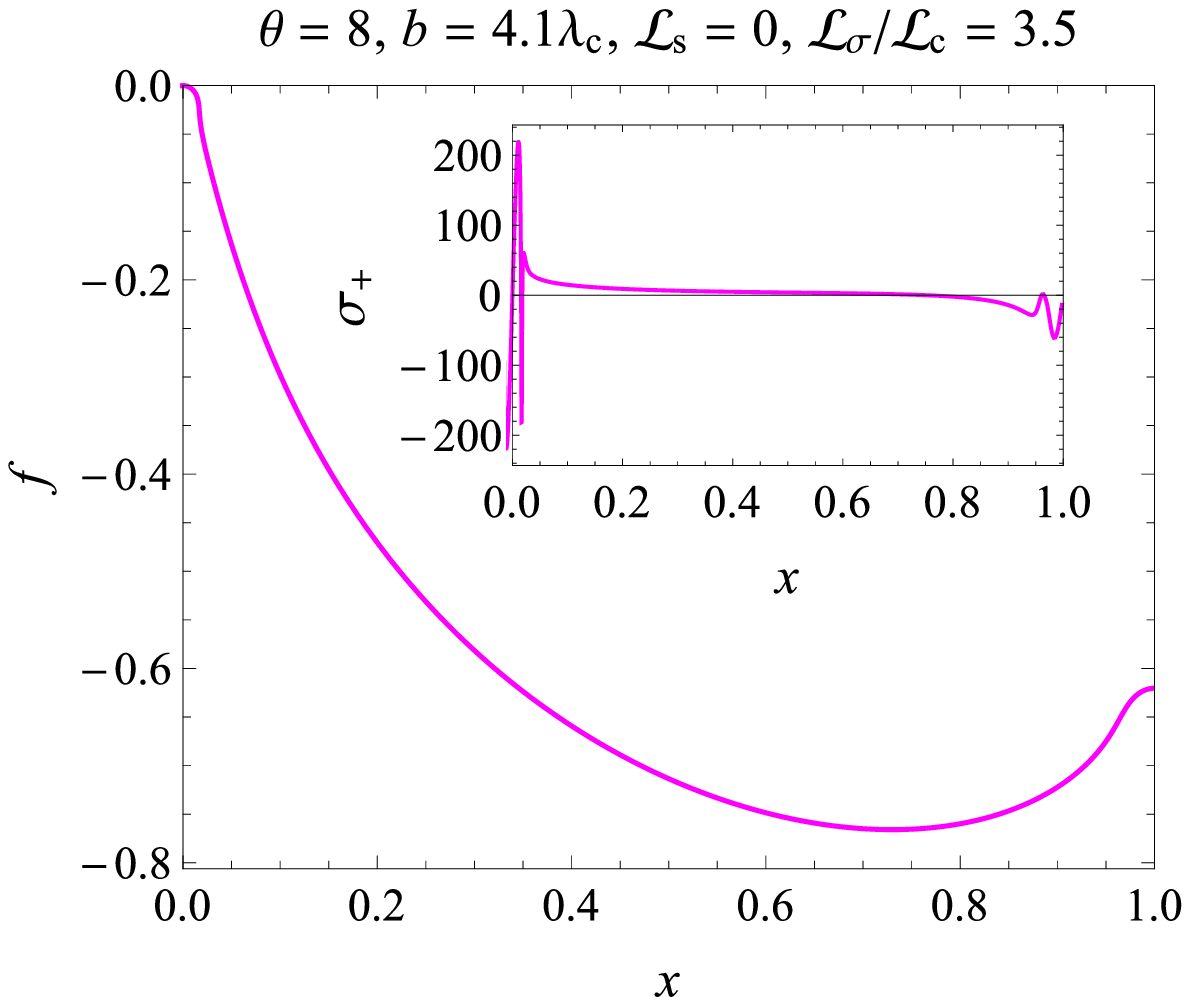}
    \caption{Front position and burnt gas vorticity distribution at the front (inset) for a flame with $\theta = 8,$ $\mathcal{L}_{\text{c}} = 0.031$ propagating in a channel of width $b=4.1\lambda_{\text{c}}$.}
    \label{fig4}
\end{figure}

Still, one might think that this breakdown of the basic assumption is not very important because it is confined to a
comparatively narrow layer near the wall. That this is not so is clearly seen from the plots of the flame propagation speed
versus $\lambda_{\text{c}}$, Fig.~\ref{fig3}. Comparison with the theoretical results (dashed lines) reveals significant
deviations from the characteristic `arch' structure, predicted by the weakly-nonlinear theory, which rapidly grow with
$\theta.$ Being proportional to the total front length, the flame speed would deviate by an amount $\sim
\mathcal{L}_{\text{c}}/b \ll 1$ if the vorticity peak affected the front shape only near the wall. Therefore, the greater
deviations exposed by Fig.~\ref{fig3} mean that what is happening near the walls actually affects the whole flame. This
conclusion is not unexpected in such an essentially nonlocal problem as the slow flame propagation. The interaction between
distant regions of the front ceases only in the case of highly elongated flames, such as those formed by a strong gravity
\cite{kazakov3,kazakov4}. It can be added that for still larger values of the gas expansion parameter, the burnt gas vorticity significantly exceeds unity also at distances $\sim b$ away from the walls, that is, virtually all over the flame front except near the vorticity nodes. In the case shown in Fig.~\ref{fig4}, for instance, the vorticity magnitude averaged over the front is approximately $11$.

\begin{figure}[ht]
    \includegraphics[width=8cm]{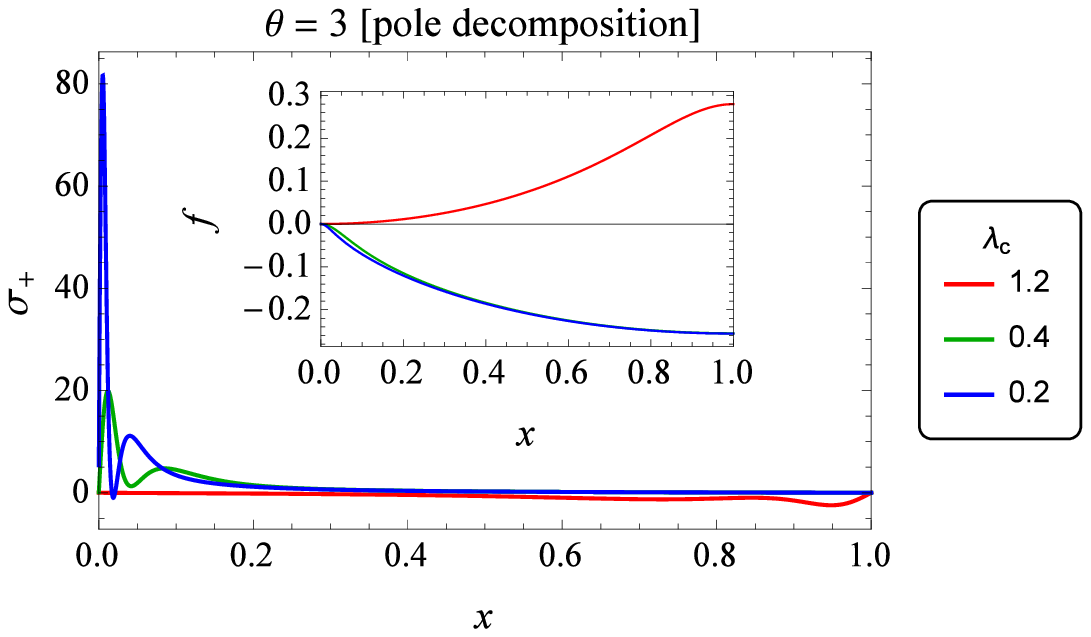}
    \caption{Burnt gas vorticity at the front as given by the pole solutions \eqref{anzats}--\eqref{solution2} of Eq.~\eqref{3ordereqf}
    for flames with $\theta=3$ and various cutoff wavelengths: $\lambda_{\text{c}}=1.2$ (red line), $\lambda_{\text{c}}=0.4$ (green line),
    and $\lambda_{\text{c}}=0.2$ (blue line). Inset: the corresponding front positions.}
    \label{fig5}
\end{figure}

Curiously, the pole solutions to the Sivashinsky--Clavin equation, when applied to flames with $\theta - 1 \gtrsim 1,$
qualitatively correctly reproduce the singular behavior of vorticity near one of the walls, if it is
calculated by substituting Eqs.~\eqref{evolutionEquation}, \eqref{vtau} into Eq.~\eqref{sigmaJump}, see Fig.~\ref{fig5}. Of course, the small-scale structure occurs in this case because the small length $\mathcal{L}_{\text{c}}$ is explicitly present in the expression for the vorticity. The fact that the
weakly-nonlinear theory predicts this behavior of vorticity, but fails at the same time to take into account its effect on the
flame structure is an indirect but quite a vivid demonstration of impossibility to apply this theory to flames with arbitrary
$\theta$ (a more direct demonstration can be found in Ref.~\cite{kazakov6}).

\begin{figure}[ht]
    \includegraphics[width=10cm]{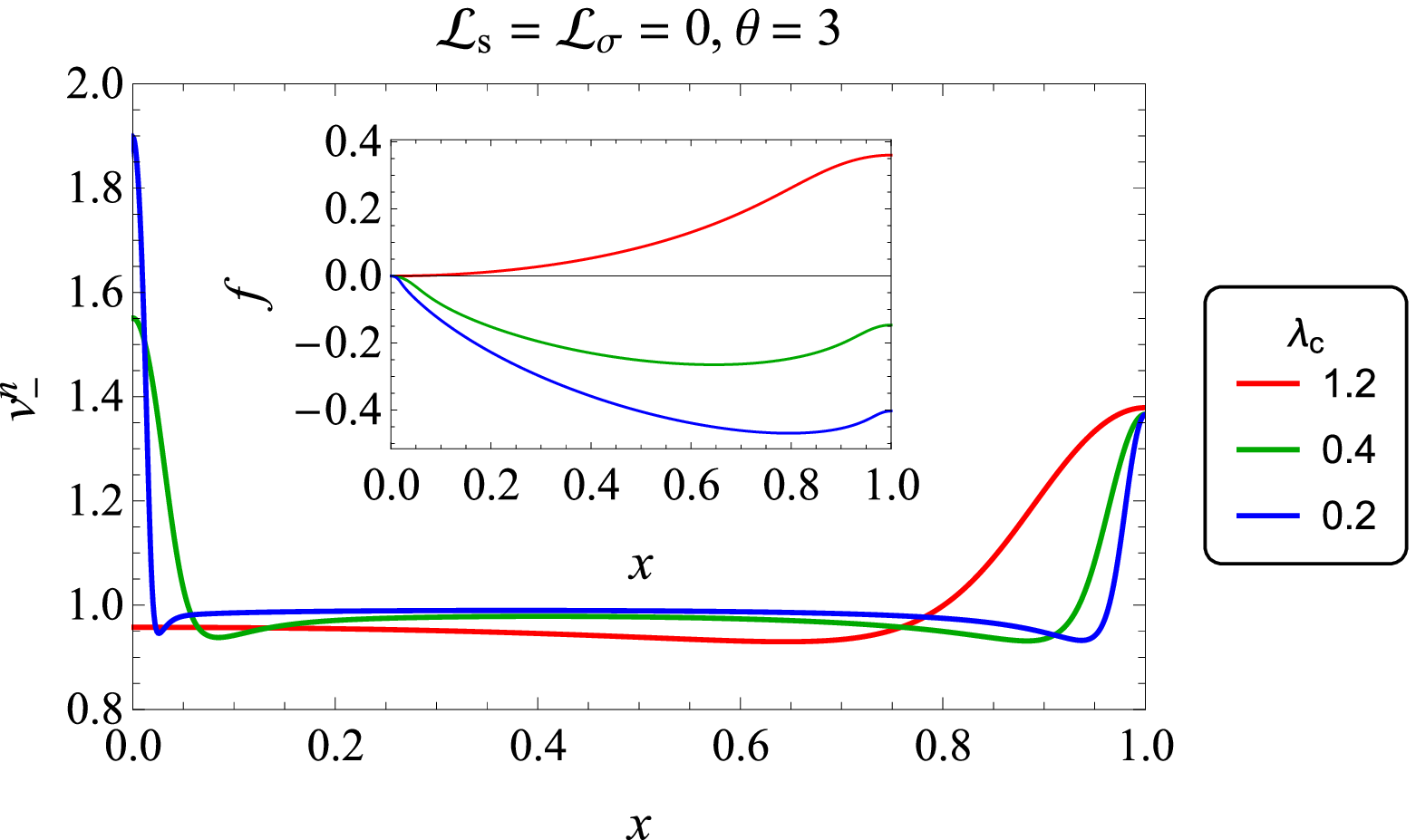}
    \caption{Normal speed of flames with $\lambda_{\text{c}}=1.2$ ($\mathcal{L}_{\text{c}} = 0.064$, red line),
    $\lambda_{\text{c}}=0.4$ ($\mathcal{L}_{\text{c}} = 0.021$, green line), and
    $\lambda_{\text{c}}=0.2$ ($\mathcal{L}_{\text{c}} = 0.011$, blue line).
    Inset: the corresponding front positions. In all the plots, $\theta = 3$ and $\mathcal{L}_{\text{s}} = \mathcal{L}_{\sigma} = 0$.}
    \label{fig6}
\end{figure}

It should be noted that not only peculiarities in the vorticity production, but also major manifestations of the
finite-front-thickness effects are confined to the near-wall layers. This is well illustrated by the normal flame speed plots
in Fig.~\ref{fig6}. The normal speed is seen to be remarkably constant across the channel except at distances
$O(\mathcal{L}_{\text{c}})$ from the walls, and the more close it is to unity, the less is $\mathcal{L}_{\text{c}}$. Thus,
nonlinear stabilization of flames with non-small $\alpha$ acts so as to reduce as much as possible the flame front curvature
and flow strain in the bulk at the cost of their enhancement near the front ends (this conclusion holds true on inclusion of
the flame stretch and compression terms).

For generic values of the Markstein parameters, enhanced vorticity production near the walls ceases only for sufficiently small
$\alpha.$ As our results show, the position and height of the two rightmost arches of the flame speed curves (which correspond
to $b\lesssim 2.5\lambda_{\text{c}}$) are described by the pole solutions reasonably well for $\alpha \lesssim 0.5$.
Restriction $\theta \lesssim 1.5$ is thus a practical limitation of applicability of the weakly-nonlinear theory in narrow
channels. Yet, even for such $\theta$'s, deviations from the predictions of this theory grow rapidly with the channel width
starting from $b\approx 3\lambda_{\text{c}},$ as further discussed in Sec.~\ref{speedrise}.

\subsection{The flame stretch effect}

It is common for theoretical as well as experimental flame studies not to distinguish the flame curvature and stretch
contributions to the normal flame speed, despite the fact that in general the two effects are quite different. There are
several reasons for that. One is that the two contributions take on the same functional form for nearly planar flames. In the
weakly-nonlinear theory, the corresponding Markstein lengths enter the equation for the front position through a single
parameter, the cutoff wavelength [cf.~Eqs.~(\ref{3ordereqf}), (\ref{lambda})]. It is thus often believed that specifying the
value of this parameter is sufficient for a global flame description. Another reason is the well-known result
\cite{matalon1982} of asymptotic analysis of the inner flame structure that in the simplest case of a one-step reaction with
high activation energy, only the stretch contributes to the normal flame speed when the discontinuity surface replacing the
front is identified with the reaction zone. Only recently was it understood \cite{kazakov5} that such an identification is not
self-consistent, and theoretical investigation of more realistic models has begun \cite{clavin2011}. On the experimental side,
accuracy of the standard flame speed measurements usually does not allow separation of the stretch and curvature contributions,
so that one length parameter turns out to be sufficient to fit the experimental data.

\begin{figure}[ht]
     $\begin{array}{cc}
            \includegraphics[height=6cm]{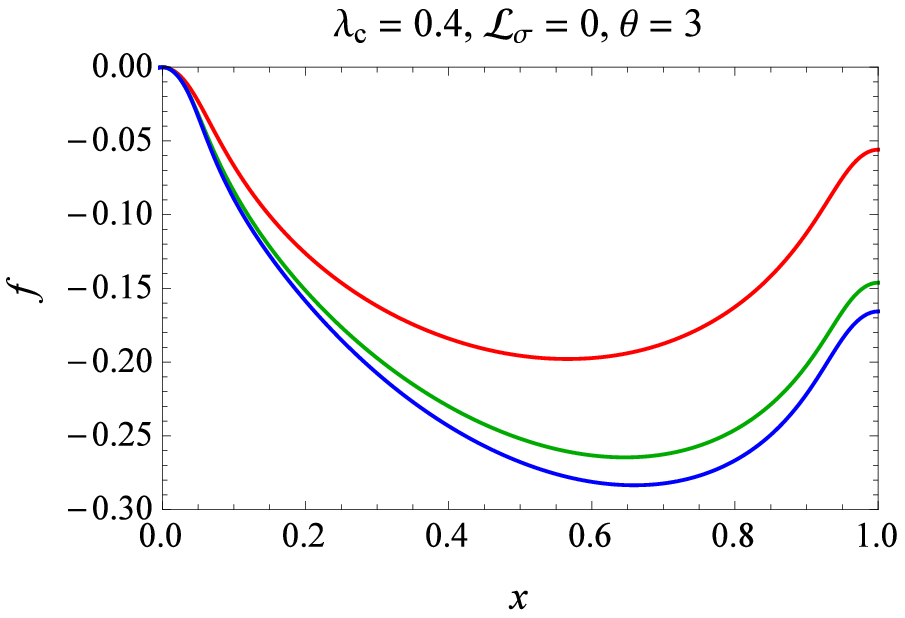} &
            \includegraphics[height=6cm]{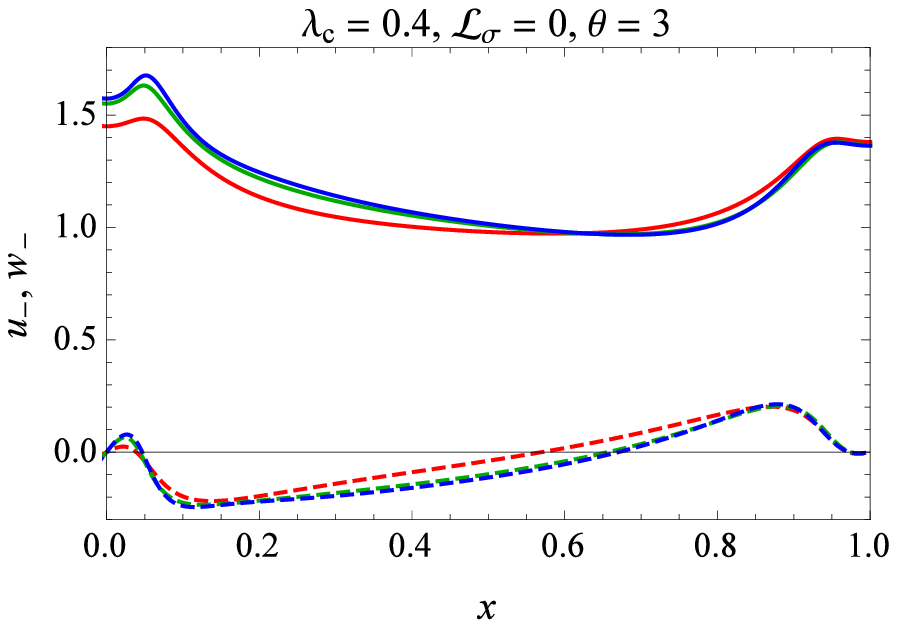} \\ \text{(a)} & \text{(b)}\\
            \includegraphics[height=6cm]{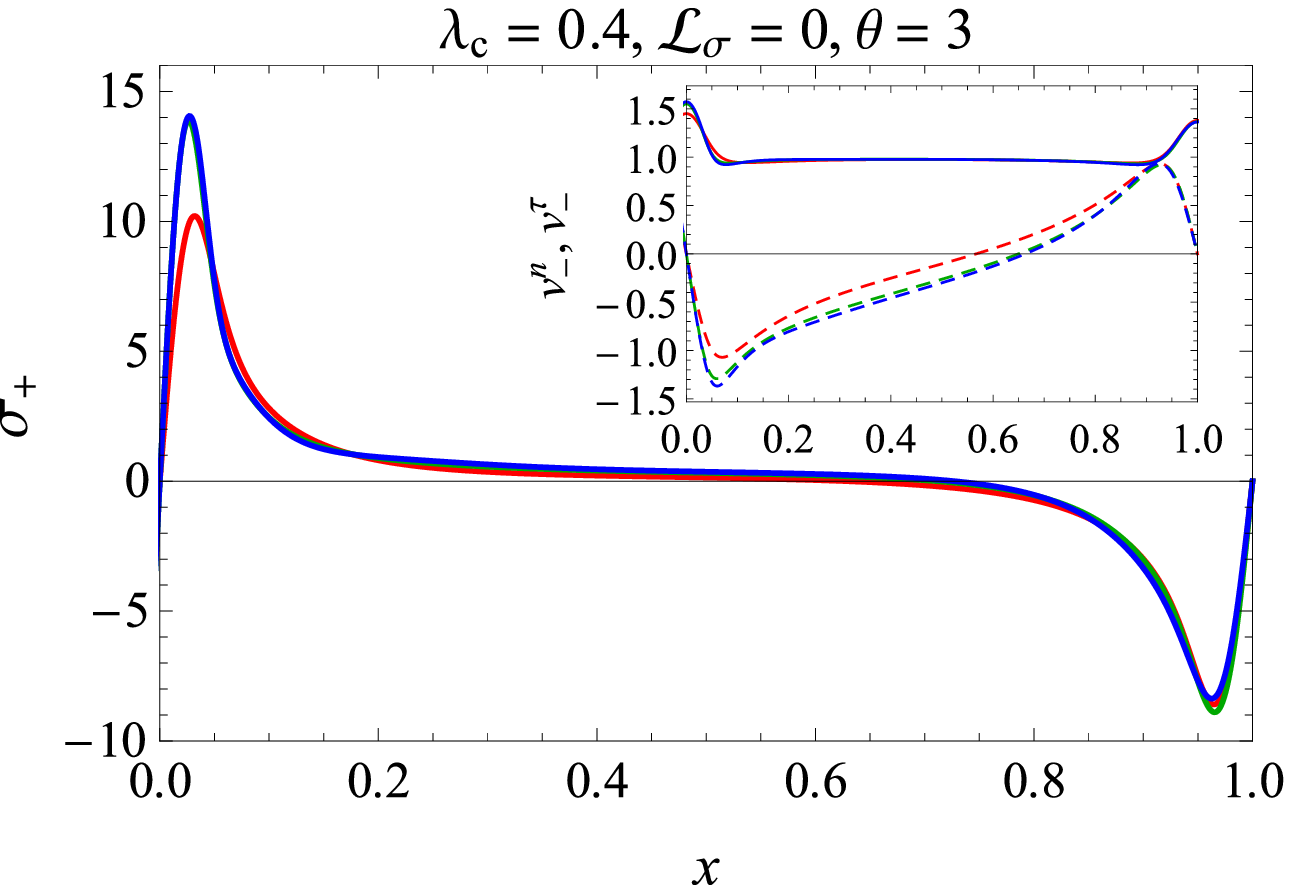} &
            \includegraphics[height=6cm]{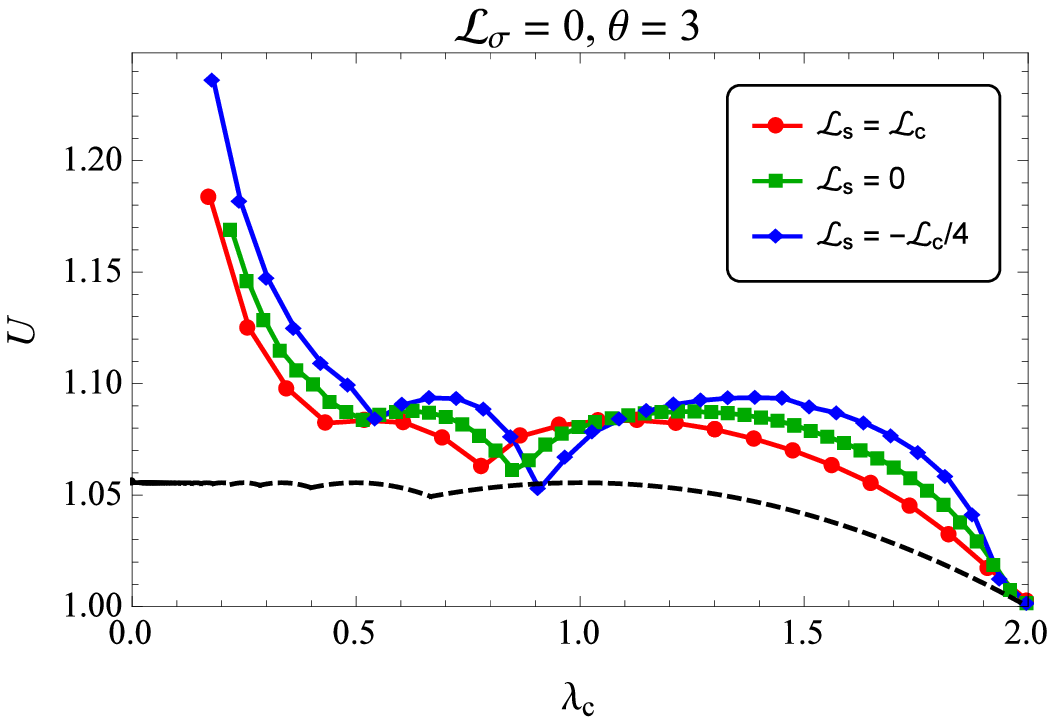} \\
             \text{(c)} & \text{(d)}
     \end{array}$
          \caption{Flame front position (a), gas velocities $u_-$ (solid lines), $w_-$ (dashed lines) (b), and burnt gas vorticity (c) for $\lambda_{\text{c}}=0.4,\theta = 3$ and $\mathcal{L}_{\text{s}} = \mathcal{L}_{\text{c}}$ (red lines), $\mathcal{L}_{\text{s}} = 0$ (green lines), $\mathcal{L}_{\text{s}} = - \mathcal{L}_{\text{c}}/4$ (blue lines); inset: the corresponding normal (solid lines) and tangential (dashed lines) fresh gas velocities. (d) Flame propagation speed versus $\lambda_{\text{c}}$ for the same $\mathcal{L}_{\text{s}}$'s.}
    \label{fig7}
\end{figure}

\begin{figure}[ht]
     $\begin{array}{cc}
            \includegraphics[height=6cm]{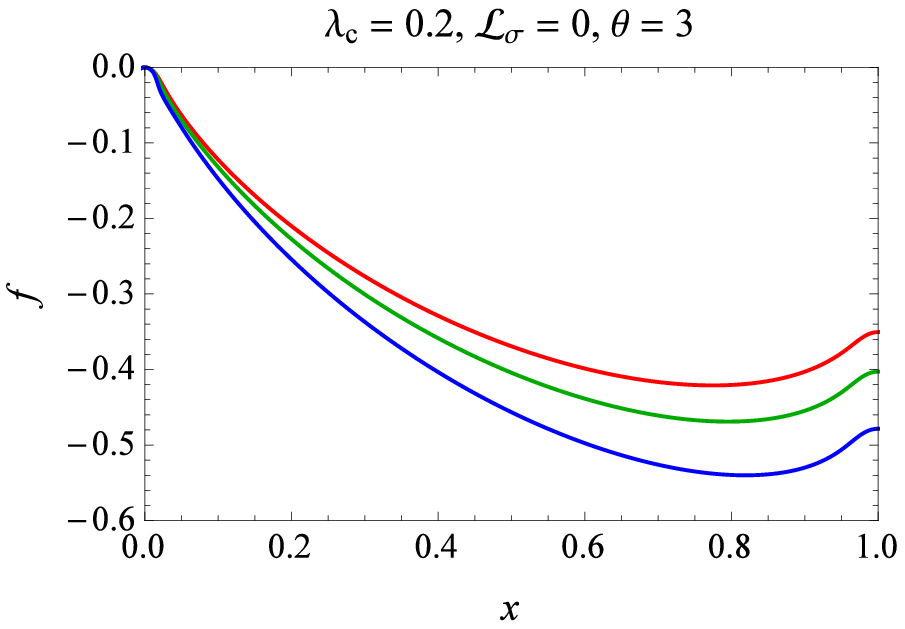} &
            \includegraphics[height=6cm]{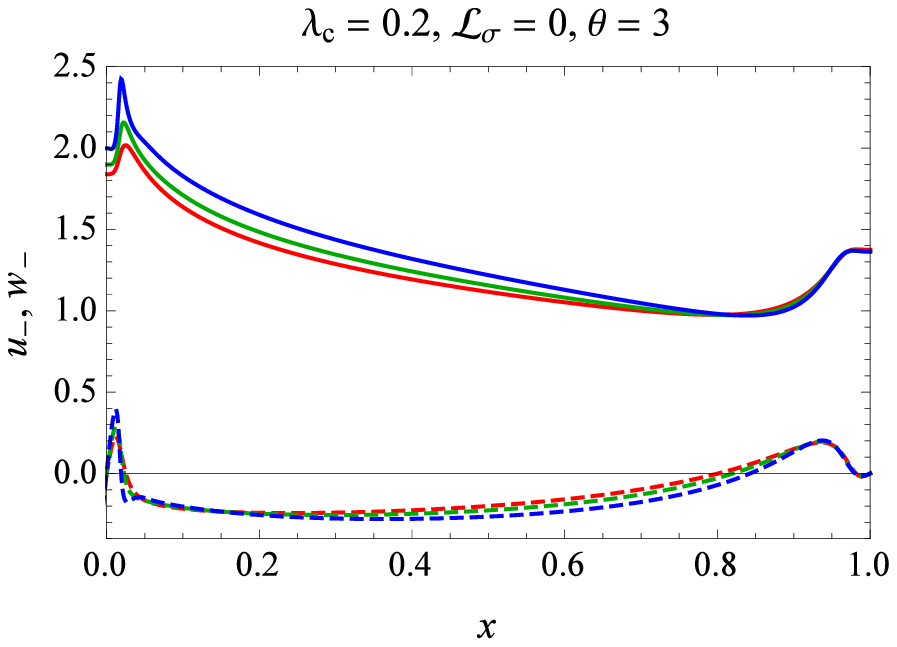} \\ \text{(a)} & \text{(b)}
     \end{array}$
     $\begin{array}{c}
     \includegraphics[height=6cm]{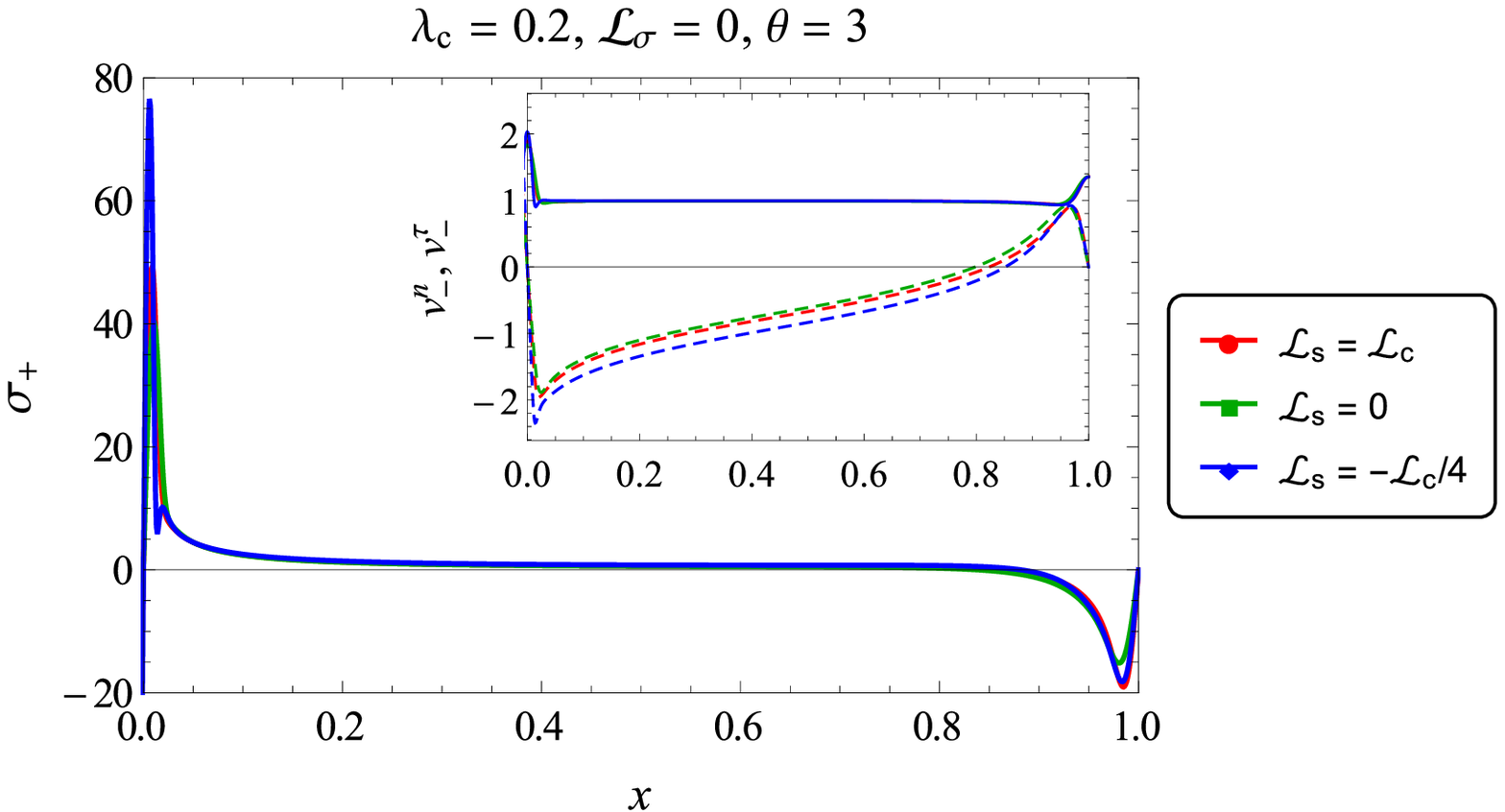}
             \\ \text{(c)}
     \end{array}$
          \caption{Same as in Figs.~\ref{fig7}(a)--\ref{fig7}(c), but for $\lambda_{\text{c}}=0.2$.}
    \label{fig8}
\end{figure}

Our results confirm that varying the ratio of the parameters $\mathcal{L}_{\text{c}},\mathcal{L}_{\text{s}}$ keeping
$\lambda_{\text{c}}$ fixed does not significantly change the flame structure. Examples are given in
Figs.~\ref{fig7}(a)--\ref{fig7}(c),~\ref{fig8} which contain plots of the flame front position, on-shell gas
velocity and burnt gas vorticity for three different pairs $\mathcal{L}_{\text{c}},\mathcal{L}_{\text{s}}$, all having the same
$\lambda_{\text{c}}.$ Still, there is an important trend produced by the flame stretch and revealed by the flame speed curves
in Fig.~\ref{fig7}(d), namely, that the flame propagation speed increases as $\mathcal{L}_{\text{s}}$ decreases, except
near the dips on the curves ({\it e.g.}, $\lambda_{\text{c}}\approx 0.9$). In this regard, things are the same as in
gravity-driven flame propagation [according to the experimental evidence \cite{levy1965} and calculations \cite{kazakov5},
propane-air flames ($\mathcal{L}_{\text{s}} >0$) propagate somewhat slower in the lean limit than methane-air flames
($\mathcal{L}_{\text{s}} < 0$)].

\subsection{The flame compression effect}

Unlike the finite-front-thickness corrections to the normal flame speed, the effect of flame compression on the flame structure
remains virtually unexplored. This is because the normal flame speed is directly measurable, and the result
(\ref{evolutionEquation_S}) of calculations can be readily compared with experiment. It is also important that similar
verifications can be made in situations where the weakly-nonlinear theory applies, such as those found in studies of various
instabilities of nearly planar flames \cite{markstein1951,barenblatt1962,searby1991}. In contrast, the flame compression
manifests itself in the gas-pressure jump at the front, hence affects only vorticity production by the flame, but not the
gas-velocity jumps themselves. In the weakly-nonlinear Sivashinsky theory \cite{siv}, the burnt gas vorticity is neglected
completely, while in the next approximation \cite{sivclav} it is considered small. As a result, in theories based on the small
$(\theta-1)$ expansion, all the three Markstein lengths $\mathcal{L}_{\text{c}},\mathcal{L}_{\text{s}}$, and
$\mathcal{L}_\sigma$ combine into a single parameter -- the cutoff wavelength of unstable flame perturbations,
$\lambda_{\text{c}}$ [cf.~Eq.~\eqref{lambda} in Appendix]. Thus, although the flame compression does change
$\lambda_{\text{c}}$, its effect is masked by those of the flame stretch and curvature. For this reason, any qualitatively new
effect associated with the flame compression will be essentially non-perturbative. As the obtained numerical solutions show,
this is the case indeed. The flame compression turns out to have a rather nontrivial impact on the structure of steady flames
with finite $\theta,$ which is not captured by the small $(\theta-1)$ expansion and is closely related to the enhanced
vorticity production near the channel walls, discussed in Sec.~\ref{gasexpansion}.

\begin{figure}[ht]
    \includegraphics[height=6cm]{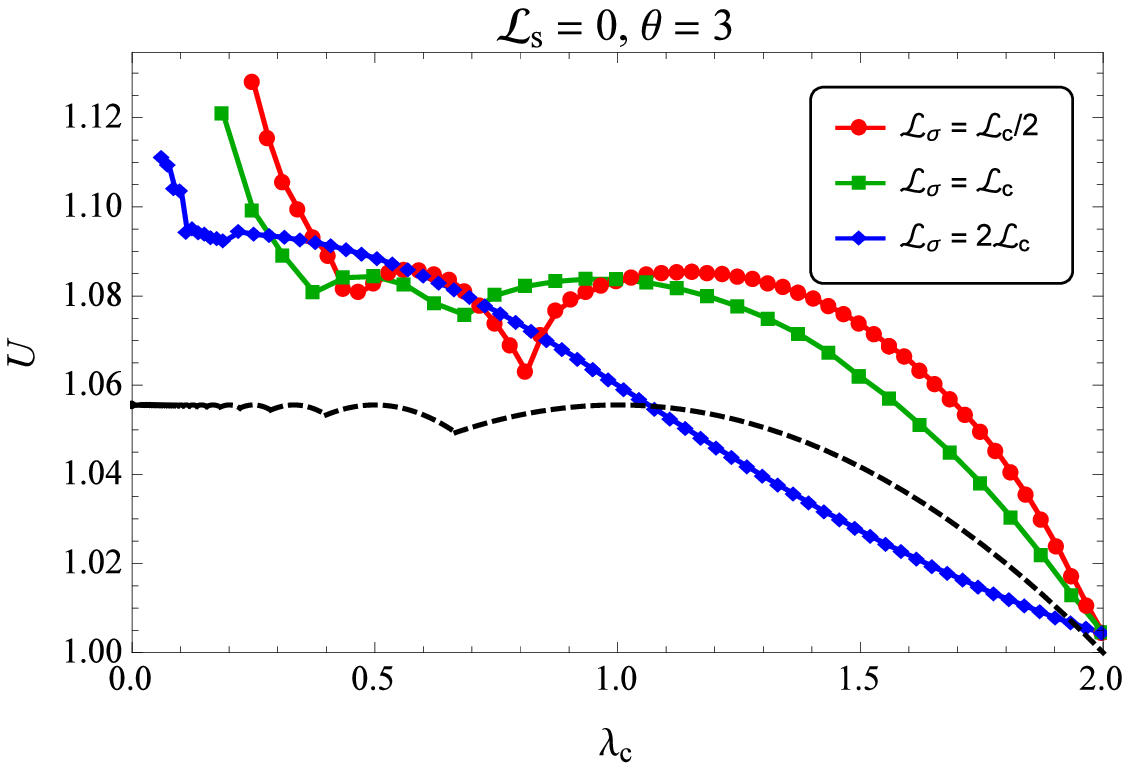}
    \caption{Flame speed versus the cutoff wavelength $\lambda_{\text{c}}$ for
    $\theta = 3$ and various $\mathcal{L}_{\sigma} / \mathcal{L}_{\text{c}}$ ratios ($\mathcal{L}_{\text{s}} = 0$ throughout).
    Red, green, and blue curves correspond to $\mathcal{L}_{\sigma} / \mathcal{L}_{\text{c}} = 0.5, 1, 2$, respectively
    (the green curve realizes condition \eqref{magic}). Dashed curve represents solutions of the Sivashinsky--Clavin equation.}
    \label{fig9}
\end{figure}

\begin{figure}[ht]
    \includegraphics[width=8cm]{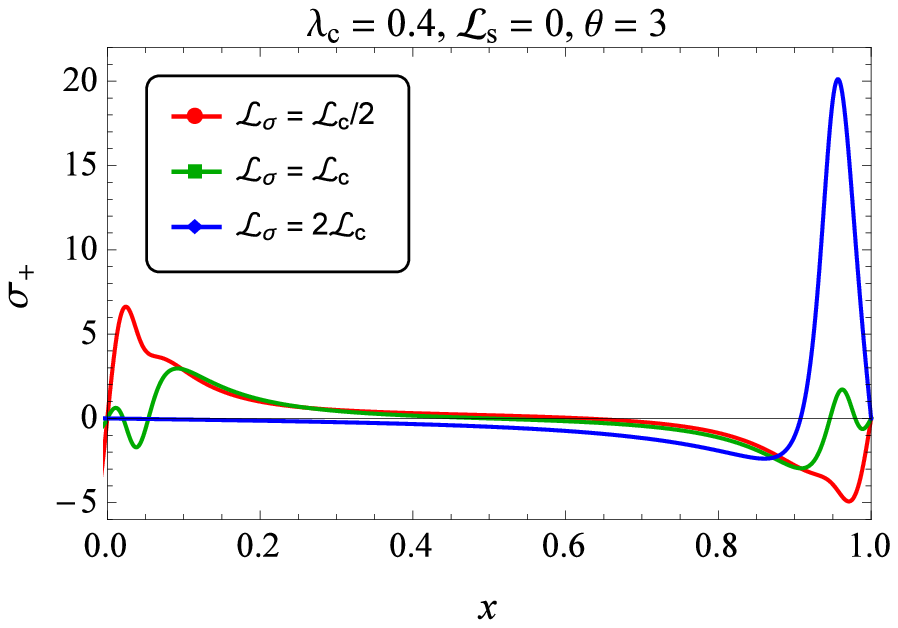}
    \caption{On-shell burnt gas vorticity in flames with $\theta=3,$ $\lambda_{\text{c}}=0.4,$
    $\mathcal{L}_{\text{s}}=0$ and $\mathcal{L}_\sigma = 0.5\mathcal{L}_{\text{c}}$ (red line),
    $\mathcal{L}_\sigma = \mathcal{L}_{\text{c}}$ (green line),  $\mathcal{L}_\sigma = 2\mathcal{L}_{\text{c}}$ (blue line).}
    \label{fig10}
\end{figure}

According to Eq.~(\ref{lambda}), contribution of the flame compression to the cutoff wavelength is negative. This means that
during the initial development of Darrieus--Landau instability, flame compression additionally destabilizes the flame, hence,
strengthens the flow nonlinearity. It is therefore somewhat surprising that in a certain range of parameters, inclusion of this
effect improves applicability of the weakly-nonlinear theory. Namely, a series of the flame velocity curves in Fig.~\ref{fig9}
demonstrate that as $\mathcal{L}_\sigma$ increases from zero to values of the order of $\mathcal{L}_{\text{c}},$ solutions of
the exact system first tend to follow more closely the pole solutions. In particular, locations of the dips on the flame speed
curve shift towards the points $b/\lambda_{\text{c}} = n+1/2,$ $n=1,2,...,$ which correspond to the appearance of new poles in
the pole decomposition [cf.~Eq.~(\ref{pmax})]. However, as $\mathcal{L}_\sigma$ increases further, deviations from the
weakly-nonlinear theory begin to grow, and the flame speed versus $\lambda_{\text{c}}/b$ ultimately becomes entirely different
from the arch-shaped curve obtained using the small-$(\theta - 1)$ expansion. This behavior is readily understood once we
observe that when the Markstein lengths $\mathcal{L}_{\text{c},\sigma}$ are related by
\begin{equation}\label{magic}
    \mathcal{L}_\sigma = \frac{\alpha}{2} \mathcal{L}_\text{c},
\end{equation}
the flame compression contribution to the vorticity jump (\ref{sigmaJump}) exactly cancels the curvature contribution coming
from the term $(v^n_-)'.$ The higher-derivative terms, which are most sensitive to rapid variations of the front slope, thus
disappear from $\sigma_+.$ As a consequence, the near-wall peaks in the vorticity production become smoothed, and the agreement
with the weakly-nonlinear theory improves. This is illustrated by the vorticity plots in Fig.~\ref{fig10}. It is seen that
inclusion of the flame compression with $\mathcal{L}_\sigma = 0.5\mathcal{L}_{\text{c}}$ reduces the vorticity peak near the
wall $x=0$ from $\sigma_+ \approx 15$ [Cf.~Fig.~\ref{fig7}(c)] to $\sigma_+ \approx 6.$ It is further damped and split into
two peaks of opposite signs with $|\sigma_+| \approx 1$ in the case $\mathcal{L}_\sigma = \mathcal{L}_{\text{c}}$ that fulfills
condition (\ref{magic}). The integral effect of the vorticity production near the wall on the global flame structure is also
reduced by the peak alternation. At last, a still larger $\mathcal{L}_\sigma$ leads to the appearance of a positive vorticity
peak $\sigma_+ \approx 22$ near the wall $x=1$ in place of the negative peak $\sigma_+ \approx - 9$ that existed for
$\mathcal{L}_\sigma = 0.$

\subsection{The flame speed rise and the role of noise}\label{speedrise}

The theory based on the weak-nonlinearity assumption predicts that the steady flame propagation speed tends to a constant value
as the channel width increases [cf.~Eqs.~(\ref{solution4}), (\ref{wmax}) and Fig.~\ref{fig14} of Appendix]. However, it is
well known from the experiment that the flame speed strongly increases with $b.$ At the same time, any laboratory flame can be
considered steady only with a certain degree of accuracy, because of various uncontrollable factors such as small mixture
inhomogeneities or turbulent motions developing in the fast burnt gas outflow. Similarly, DNS show a strong rise of the flame
speed starting from some critical $b,$ but it is difficult to obtain a truly steady flame in sufficiently wide computational
domains: the errors caused by the numerical roundoff, finiteness of the domain lateral extent, \etc produce small perturbations
of the flame which can grow large before they are carried away downstream, and thus lead to a noticeable increase in the flame
speed or even trigger transitions between different propagation regimes. It was, therefore, suggested that these flow
irregularities, or noise, are the reason for the discrepancy between the theory and observations
\cite{procaccia1997,procaccia1998,almarcha2015}. Specifically, it was shown in Refs.~\cite{procaccia1997,procaccia1998} that by
introducing a noise term into the Sivashinsky equation and choosing appropriately the noise spectrum and intensity, one can
obtain the flame speed scaling with the channel width as $U \sim b^{\mu},$ where $\mu$ can take on different values, {\it
e.g.}, $0.35,0.42$, and even $\mu > 1.5$.

It is true that the noise is present in the real experiment as well as in simulations, and there is little doubt that the flame
speed can be raised to the observed values by taking the noise magnitude sufficiently large. But before declaring the noise
responsible for the found discrepancies, it is worth exploring the possibility that the flame speed rise is a steady-flame
phenomenon which is essentially non-perturbative, in that it is not captured by the weak-nonlinearity treatment. The master
equation (\ref{masterEquation}) is a perfect means for this purpose: being time-independent by construction, it is free of the
unsteady phenomena caused by the noise, and being an exact consequence of the fundamental gasdynamic equations, it takes full
account of the flow nonlinearities generated by a steady flame.

In the flame speed plots presented above, the speed is seen to steeply rise to the left of the second dip, which in the
pole-decomposition picture corresponds to $b = (2+1/2)\lambda_{\text{c}}.$ To see how this compares with the known behavior of unsteady flames, we use the DNS results of Ref.~\cite{Liberman}. To identify the region in the parameter space that corresponds
to the conditions of Ref.~\cite{Liberman} we take into account that these DNS were carried out in the case of unit Lewis
number, Prandtl number equal to 0.5, temperature-independent thermal conductivity, and the reduced activation energy of a
one-step chemical reaction equal to 7. The latter comparatively large value suggests to adopt the thin reaction zone
approximation, within which the Markstein lengths for the specified values of Lewis and Prandtl numbers read (explicit
expressions given in Ref.~\cite{class2003} are valid for the gas thermal conductivity scaling as the square root of
temperature; in the case of interest, expressions (\ref{infinite}) can be obtained from the general formulas (3.32)--(4.20) of
Ref.~\cite{class2003})
\begin{equation}\label{infinite}
    \mathcal{L}_{\text{c}} = \frac{\theta \ln\theta}{\theta - 1}l_{\text{f}}, \quad \mathcal{L}_{\text{s}} = 0, \quad
    \mathcal{L}_\sigma = \frac{\theta \ln\theta}{2}l_{\text{f}}.
\end{equation}

\begin{figure}[ht]
    \includegraphics[width=10cm]{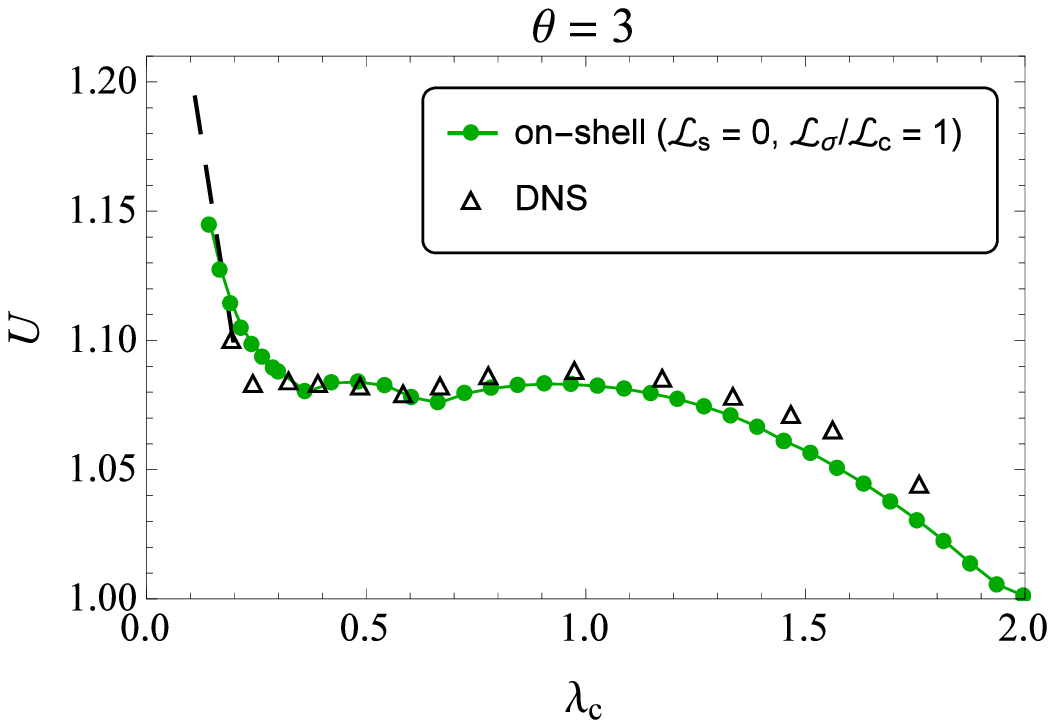}
    \caption{Propagation speed of flames with $\theta=3$ versus the cutoff wavelength as given by the on-shell solutions
    with Markstein lengths (\ref{infinite}) (circles), and by DNS \cite{Liberman} (triangles).}
    \label{fig11}
\end{figure}

\begin{figure}[ht]
    \includegraphics[width=10cm]{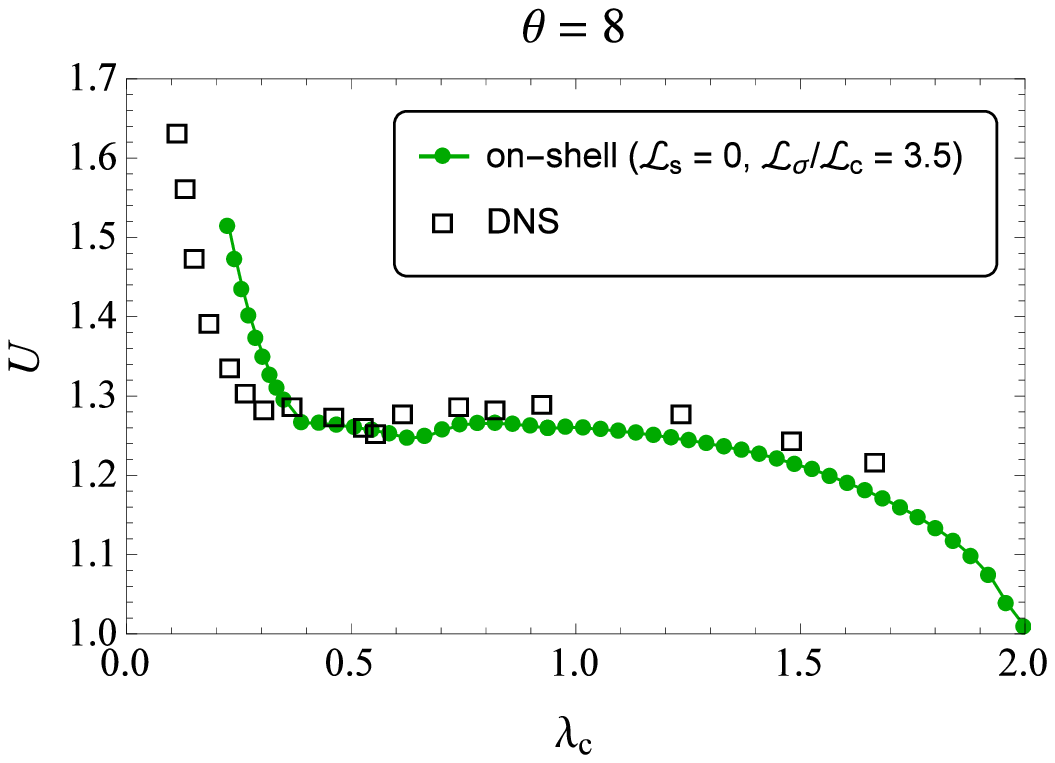}
    \caption{Same as in Fig.~\ref{fig11}, but for $\theta=8$.}
    \label{fig12}
\end{figure}

It is easy to check a curious fact that these lengths fulfill condition (\ref{magic}). Our results for the flame speed versus
cutoff wavelength are compared with the DNS data in Figs.~\ref{fig11}, \ref{fig12}. Although the values (\ref{infinite}) pertain to the limit of infinite activation energy, so that agreement with the DNS can be improved by varying Markstein parameters around these values,\footnote{Increasing $\mathcal{L}_\sigma$ shifts the flame speed curves leftwards, while $\mathcal{L}_\text{s}$ controls their height (Cf. Figs.~\ref{fig7}(d), \ref{fig9}). But it is also possible that the slight horizontal shift between our curves and the DNS data, which is evident in Figs.~\ref{fig11}, \ref{fig12} from the location of the dips, is due to a difference in measuring the cutoff wavelength: we define $\lambda_{\text{c}}$ as twice the channel width at which the flame speed reaches the value $(1+\varepsilon)$ with $\varepsilon \ll 1$ (usually $\varepsilon = 0.005$), whereas in Ref.~\cite{Liberman} it is inferred from the dispersion of the perturbation growth rate during the linear stage of Darrieus--Landau instability (O. Peil, private communication).} the comparison leaves no doubt that the two datasets describe the same flame-speed behavior, in particular, the same phenomenon of the flame speed rise at small $\lambda_{\text{c}}$. The piece of the DNS data describing this rise is shown in Fig.~\ref{fig11} as a
dashed line because of a significant scatter in the data caused by numerical noise.\footnote{O.~Peil, private communication.}
Since there is no such problem with the steady on-shell solutions, we conclude that {\it the speed rise at small
$\lambda_{\text{c}}$ is not related to noise or some flame instability, but is an intrinsic property of steady flames.}

\begin{figure}[ht]
    \includegraphics[width=10cm]{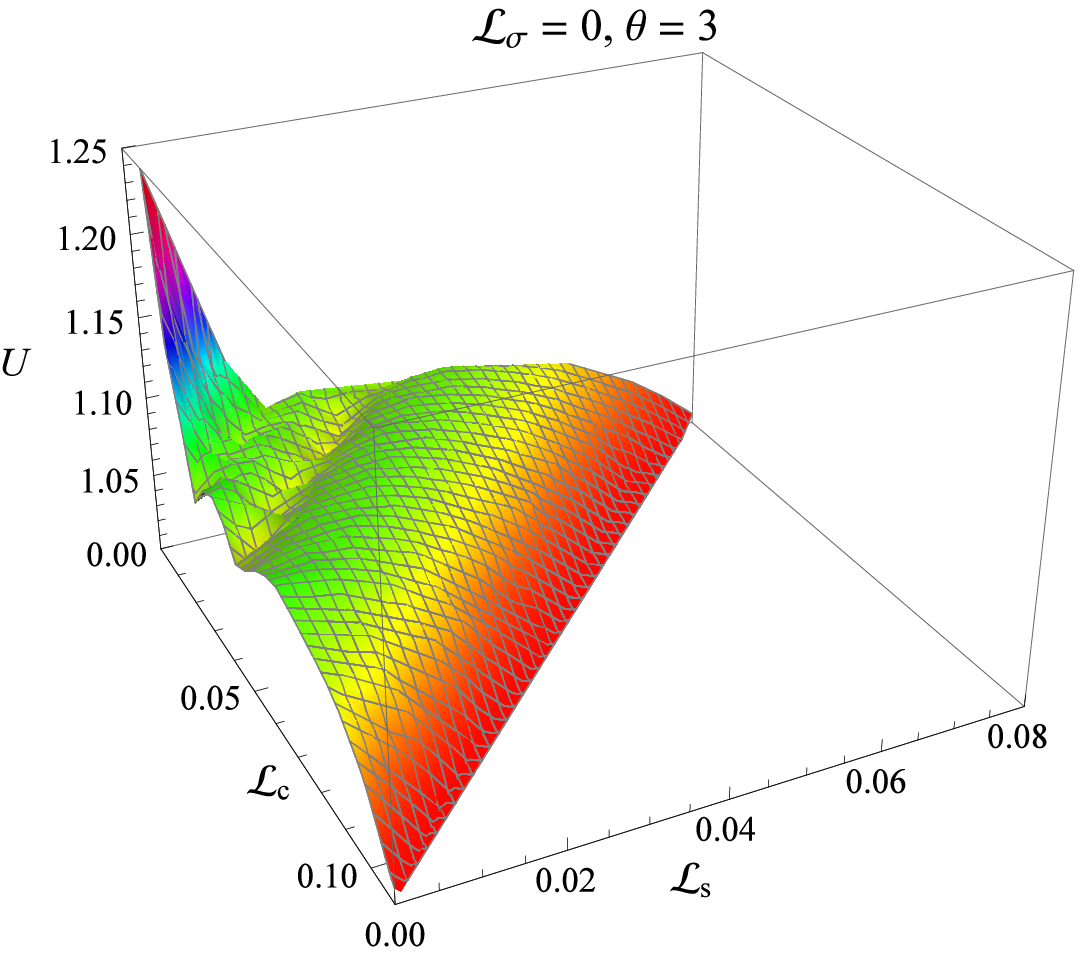}
    \caption{Propagation speed of flames with $\theta=3$ as a function of $\mathcal{L}_{\text{c}},
    \mathcal{L}_{\text{s}}$.}
    \label{fig13}
\end{figure}

To demonstrate that the found flame speed rise is quite universal, not just specific to the particular choice (\ref{infinite})
of Markstein lengths, we present in Fig.~\ref{fig13} a three-dimensional plot of the flame speed as a function of two
independent parameters $\mathcal{L}_{\text{c}}$ and $\mathcal{L}_{\text{s}}.$ For sufficiently  small values of
$\mathcal{L}_{\sigma}$ [such as the one given by Eq.~\eqref{magic}, or less], the threshold of the speed rise is $b\approx
2\lambda_{\text{c}} \div 3\lambda_{\text{c}},$ but it noticeably grows with $\mathcal{L}_{\sigma}$ [Cf. the case $\mathcal{L}_{\sigma} = 2\mathcal{L}_{\text{c}}$ in Fig.\ref{fig9} where it is $\approx
12\lambda_{\text{c}}$]. On the other hand, the threshold value turns out to be much less sensitive to the flame compression
when expressed in terms of the curvature and stretch Markstein lengths. For $\mathcal{L}_{\text{s}}=0,$ for instance, it is $b
\approx 35\mathcal{L}_{\text{c}}$ almost regardless of $\mathcal{L}_{\sigma}$.

\section{Conclusions}\label{sec:Conclusion}

The on-shell description of flame propagation makes it unnecessary to resolve the flow structure in the streamwise direction.
This dimensional reduction of the problem gives a great advantage over conventional methods based on direct solution of the
fundamental gasdynamic equations. We have shown that the system of the on-shell equations for a steady flame can be solved
numerically in a comparatively simple way using two-level nested fixed-point iterations. Inner iterations give a solution of
the master equation for a fixed, current flame front position, which is then used to construct a subsequent outer iteration of
the flame front itself -- namely, to solve the evolution equation for the obtained gas velocity distribution along it.

Another important advantage over experimental studies and DNS is that our approach permits separate investigation of various
effects controlling the flame evolution. The results of such an investigation of the channel flame propagation, carried out in
Sec.~\ref{sec:Results}, lead us to the following main conclusions:
\begin{itemize}
  \item As a result of nonlinear flame stabilization in a channel, layers near the channel walls develop, which are characterized
        by an enhanced vorticity production and large flame front curvature. Markstein length $\mathcal{L}_{\text{c}}$ is the
        characteristic length of the flow fields in these layers, wherein the magnitudes of the vorticity and front curvature
        strongly grow with the gas expansion coefficient $\theta$. The existence of these layers makes the weakly-nonlinear theory
        inapplicable to flames with $\theta - 1 \gtrsim 1$ even in narrow channels.
  \item Flame compression significantly affects the flame structure by modifying the vorticity production in the flame. This modification
        is especially pronounced in the same near-wall layers in which rapid variations of the flow variables take place. In particular, the
        flame compression violently alters the dependence of the flame propagation speed on the channel width, so that the usual `arch'-shaped
        profile predicted by the weakly-nonlinear theory turns out to be smoothed and nearly monotonic for sufficiently large $\mathcal{L}_{\sigma}$.
  \item A steep rise of the flame propagation speed observed in sufficiently wide channels is a steady flame phenomenon up to at least
        $b = 200\mathcal{L}_{\text{c}}$. It has nothing to do with noise or any other possible flame unsteadiness that can be present in DNS
        or an actual experiment.
\end{itemize}

These observations in turn raise an important and quite nontrivial question of an ultimate structure of the near-wall layers in
the limit $b/\lambda_{\text{c}} \to \infty$. Since the magnitude of the vorticity peaks grows (apparently, without bound) as
$b$ increases, it is clear that either the inner flame structure or the structure of the gas flow in these layers must
eventually change so as to keep the vorticity bounded. Our results show that as $b$ increases, the front slope becomes very
large just before it drops down to zero at the channel wall, that is, the front turns out to be nearly parallel to the wall. This
suggests that the required change in the inner flame structure might be its local quenching. Formulated as the vanishing normal
velocity $v^n_-$ at the wall, this would also align the front with the wall, but eliminate the undesirable large front
curvature. Another option -- a change in the outer flow structure -- can be realized through the formation of a stagnation zone
in the burnt gas flow. This would lift condition $f' = 0$ at the wall, and hence relax the flow strain in the near-wall layers.
Hopefully, resolution of this question will help to identify the boundary conditions relevant to the flame propagation in situations where the flame propagation speed largely exceeds its normal speed.

\acknowledgments %
The authors thank H.~El-Rabii for fruitful discussions, O.~Peil  for clarification and discussion of the paper
\cite{Liberman}, S.~Matveev for his kind advice on the application of the Anderson acceleration method, and A.~Zhugayevych's group at Skolkovo Institute of Science and Technology for the help with the computational resources. The reported study was partially supported by RFBR, research project No. 13-02-91054~a, and by Supercomputing
Center of Moscow State University \cite{Lomonosov}.

\appendix

\section{Equation for the front position in the first post-Sivashinsky approximation}\label{sec:Appendix}

In this appendix, we derive an equation for the flame front position in the first post-Sivashinsky approximation (a steady
version of the Sivashinsky--Clavin equation \cite{sivclav}) taking into account the flame compression effect. This is primarily
to recapitulate the main assumptions underlying the weak-nonlinearity analysis, and to recall basic properties of the pole decomposition solutions \cite{thual1985}.

Soon after the onset of the planar flame instability, the cutoff wavelength of unstable flame perturbations, $\lambda_{\text{c}},$ becomes the smallest\footnote{As long as $\lambda_{\text{c}}<b$; however, since the planar flame is stable for $\lambda_{\text{c}}> 2b,$ the distinction between $\lambda_{\text{c}}$ and $b$ as characteristic lengths is immaterial in the case $\lambda_{\text{c}}> b$.} characteristic length of the weakly curved flame, because perturbations with wavelengths smaller than $\lambda_{\text{c}}$ rapidly die out during the linear stage of Darrieus--Landau instability. A starting point of the small-$(\theta-1)$ expansion is the fact of the linear stability analysis that $\lambda_{\text{c}} \sim l_\text{f}/(\theta - 1).$ This suggests that for a given front thickness $l_\text{f}$, any flame pattern will be ``stretched out'' in the $x$-direction as $\theta \to 1$. In other words, {\it assuming that $\lambda_{\text{c}}$ remains the smallest length during the subsequent flame evolution}, the flame nonlinearity can be made weak by choosing $(\theta -1)$ sufficiently small. $(\theta -1)\equiv \alpha$ thus becomes a small parameter of the weak-nonlinearity expansion. Since $\lambda_{\text{c}} = O(1/\alpha)$, the $x$-differentiation of an on-shell quantity raises its smallness order within this expansion by one. For instance, the front slope is to be treated as a first-order quantity, $f' = O(\alpha),$ whereas $f'' = O(\alpha^2).$ Similarly, $w' = O(\alpha w),$ $u' = O(\alpha (u-1)).$

The first post-Sivashinsky approximation corresponds to retaining terms up to $O(\alpha^4)$ in the master
equation~\eqref{masterEquation}. As is known from the Sivashinsky's theory, both $(u_- - 1)$ and $w_-$ are $O(\alpha^2)$.
Therefore, the first two terms in the braces in Eq.~(\ref{masterEquation}) are expanded as
\begin{eqnarray}
    \frac{Nv^n_+\sigma_+\omega_+}{v^2_+} &=& - \frac{\alpha}{\theta} f'f'' + \alpha S' - 2\mathcal{L}_\sigma f''' + O(\alpha^5), \\
    {[}\omega{]}'  &=& - {\rm i}\alpha f'' - \frac{\alpha}{2}\left(f'^2\right)' - \alpha S' + O(\alpha^5),
\end{eqnarray}
while the third term in the braces does not contribute within this order, because the real part of the integral is zero identically
[by virtue of the parity properties (\ref{reflect})], whereas its imaginary part is only $O(\alpha^6)$. In the above formulas,
$S$ can be replaced by its leading-order expression $S = (\mathcal{L}_{\text{c}} + \mathcal{L}_{\text{s}})f''$, since $v^{\tau}
= f' + O(\alpha^2)$. Substituting this into the master equation, separating its real and imaginary parts (the latter is needed
only up to the third order), and integrating over $x$ yields
\begin{eqnarray}
    2u_- - 2\left[\alpha(\mathcal{L}_{\text{c}} + \mathcal{L}_{\text{s}}) - \mathcal{L}_\sigma\right]f'' + \alpha \hat{H}f' = C, \nonumber\\
    2w_- - \alpha f' = 0,
\end{eqnarray}
where $C$ is an integration constant and $\hat{H}$ is the Hilbert operator (that is, $\hat{\Hilbert}$ with $f\equiv 0$). $u_-$
and $w_-$ found from these equations are to be substituted into Eq.~(\ref{evolutionEquation}) with $S$ expanded to the third
order. For this purpose, we write
\begin{eqnarray}\label{vtau}
    v^{\tau} = \frac{w_- + f' u_-}{N} = \frac{w_- + f'(f'w_- + N - SN)}{N} = f' + w_- N - f'S = f' + w_- + O(\alpha^3) = \frac{\theta + 1}{2}f' + O(\alpha^3),
\end{eqnarray}
and then
$$
    S = \mathcal{L}_{\text{c}}f'' + \mathcal{L}_{\text{s}} \frac{\theta + 1}{2}f'' + O(\alpha^4).
$$
We thus find
\begin{eqnarray}\label{3ordereqf}
    - \theta(U - 1) + \frac{\theta}{2}(f')^2 = \frac{\theta-1}{2}\left(- \hat{H}f' + \frac{\lambda_{\text{c}}}{2\pi}f''\right),
\end{eqnarray}
where the cutoff wavelength
\begin{eqnarray}\label{lambda}
    \lambda_{\text{c}} = \frac{4\pi}{\theta - 1}\left[\theta \mathcal{L}_{\text{c}} + \frac{3\theta - 1}{2}\mathcal{L}_{\text{s}} - \mathcal{L}_\sigma\right].
\end{eqnarray}
The constant $C$ has been expressed via the flame propagation speed $U$ with respect to the gas far upstream by averaging the
equation over $x\in [-1,1],$ and taking into account the periodicity of $f(x)$ and formula
$$
    U = 1 + \int_{0}^{1} dx \frac{(f')^2}{2} + O(\alpha^4).
$$
To reiterate, the assumption of smallness of the front curvature, $f'',$ is essential for the weak-nonlinearity analysis: since
all other terms in Eq.~(\ref{3ordereqf}) are $O(\alpha^2),$ so must be $f''.$

\begin{figure}[ht]
    \includegraphics[width=10cm]{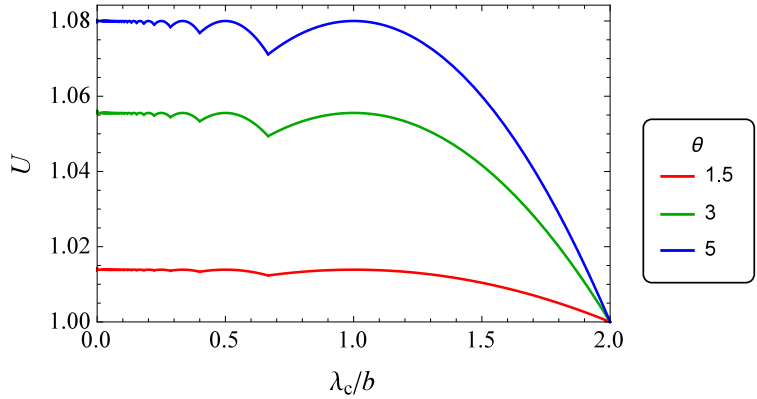}
    \caption{The flame speed versus $\lambda_{\text{c}}/b$ as given by the pole solutions of Eq.~\eqref{3ordereqf}.}
    \label{fig14}
\end{figure}

It is seen that apart from a modification of the cutoff wavelength, account of the flame compression effect does not change the
functional structure of the equation for flames with weak gas expansion. This equation is therefore readily solved using the
pole decomposition \cite{thual1985}
\begin{eqnarray}\label{anzats}
    f(x) = A \sum_{k = 1}^{2 P} \ln\sin\left[\frac{\pi}{2}(x - x_k)\right].
\end{eqnarray}
The amplitude $A$ and the complex poles $x_k,$ $k = 1,...,2P$ are found by substituting this decomposition into
Eq.~(\ref{3ordereqf}) and using the formulas
\begin{gather*}
    \hat{H}f' = - \frac{\pi A}{2}\sum_{k = 1}^{2 P}\left\{1 + \ii \sgn(\IIm x_k)\cot\left[\frac{\pi}{2}(x - x_k)\right]\right\}, \quad
    \sgn(x) \equiv \frac{x}{|x|},\\
    \cot x \cot y = -1 + \cot(x - y )(\cot y - \cot x).
\end{gather*}
This gives
\begin{eqnarray}\label{solution1}
    A &=& - \frac{\lambda_{\text{c}}}{2\pi}\frac{\theta - 1 }{\theta}\,,
    \nonumber\\ U - 1 &=& \frac{(\theta - 1)^2}{2\theta^2
    }\frac{P\lambda_{\text{c}}}{2}\left(1 - \frac{P\lambda_{\text{c}}}{2}\right),
\end{eqnarray}
and a set of equations for $x_k$
\begin{eqnarray}&&\label{solution2}
    \ii\sgn(\IIm x_k) + \frac{\lambda_{\text{c}}}{2}\sum\limits_{\genfrac{}{}{0pt}{}{m = 1}{m\ne k}}^{2 P}
    \cot\left[\frac{\pi}{2}(x_k - x_m)\right] = 0, ~k = 1,...,2P\,.
\end{eqnarray}
For each value of $\lambda_{\text{c}}$ there is a number of solutions corresponding to different numbers $P$ of complex
conjugate pole pairs. The solution with the maximal flame speed is that with the maximal number of poles, which are vertically
aligned in the complex $x$-plane. The maximal $P$ can be inferred from Eq.~(\ref{solution2}) by setting $k = k_0,$ where
$x_{k_0}$ is the uppermost. Restoring the ordinary units for clarity, one finds
\begin{eqnarray}\label{pmax}
    P_{\rm max} = {\rm Int} \left(\frac{b}{\lambda_{\text{c}}} + \frac{1}{2}\right),
\end{eqnarray}
where ${\rm Int}(x)$ denotes the integer part of $x.$ Thus, the maximal flame speed is
\begin{equation}\label{solution4}
    U_{\rm max} = U_\text{f} + 2 W_{\rm max}P_{\rm max}\frac{\lambda_{\text{c}}}{b}\left(1 - \frac{P_{\rm max}}{2}\frac{\lambda_{\text{c}}}{b}\right),
\end{equation}
where
\begin{eqnarray}\label{wmax}
    W_{\rm max} = \frac{(\theta - 1)^2}{8\theta^2}U_\text{f}\,.
\end{eqnarray}
$U_{\rm max}$ versus $\lambda_{\text{c}}/b$ is plotted in Fig.~\ref{fig14}. This plot illustrates a characteristic property
of the steady solutions to the Sivashinsky equation and its modifications derived under the weak-nonlinearity assumption,
namely, that {\it the maximal flame speed tends to a constant value $(1+W_{\rm max})$ as the channel width increases.}

\begin{thebibliography}{123}
\bibitem{note1}
        Though extension of the on-shell description to three dimensions is still unavailable, the reason to believe in its existence is the same as in two
        dimensions. The formally infinite speed of sound, implied by the flow incompressibility, means that any change in the flame structure is instantly ``felt''
        by the entire flow, and vice versa. All information about the flow at any instant can thus be thought of as imprinted in the on-shell fields at the same
        instant. On the other hand, these fields define, together with the flow conditions at the tube walls, boundary conditions for the entire gas flow.
        Therefore, the space-time history of these fields up to a given instant determines their subsequent evolution, hence provides an on-shell description of the
        flame dynamics.
\end{thebibliography}
\end{document}